\documentclass[fleqn,10pt]{wlscirep}
\usepackage[utf8]{inputenc}
\usepackage[T1]{fontenc}
\usepackage{lineno}
\usepackage{soul}
\usepackage{amsmath}
\usepackage{upgreek}
\usepackage{bm}
\usepackage{amssymb}
\usepackage{pifont}
\usepackage{multirow}
\usepackage{subfigure}
\usepackage[numbers, compress]{natbib}
\usepackage[font=footnotesize,labelfont=bf, justification=justified, format=plain]{caption}
\usepackage{makecell}
\usepackage{comment}
\usepackage{caption}
\usepackage{subcaption}
\usepackage{xurl}

\usepackage{xr}
\makeatletter
\newcommand*{\addFileDependency}[1]{
\typeout{(#1)}
\@addtofilelist{#1}
\IfFileExists{#1}{}{\typeout{No file #1.}}
}\makeatother

\newcommand*{\myexternaldocument}[1]{%
\externaldocument{#1}%
\addFileDependency{#1.tex}%
\addFileDependency{#1.aux}%
}
\myexternaldocument{supplementary}

\title{Implementing Trust in Non-Small Cell Lung Cancer Diagnosis with a Conformalized Uncertainty-Aware AI Framework in Whole-Slide Images}

\author[1,$\dag$, $*$]{Xiaoge Zhang}
\author[1,$\dag$]{Tao Wang}
\author[2,$\dag$]{Chao Yan}
\author[3]{Fedaa Najdawi}
\author[4]{Kai Zhou}
\author[5]{Yuan Ma}
\author[6]{Yiu-ming Cheung}
\author[2, 7, 8, $*$]{Bradley A. Malin}

\affil[1]{Department of Industrial and Systems Engineering, The Hong Kong Polytechnic University, Kowloon, Hong Kong}
\affil[2]{Department of Biomedical Informatics, Vanderbilt University Medical Center, Nashville, TN, USA}
\affil[3]{Department of Pathology, Microbiology and Immunology, Vanderbilt University Medical Center, Nashville, TN, USA}
\affil[4]{Department of Computing, The Hong Kong Polytechnic University, Kowloon, Hong Kong}
\affil[5]{Department of Mechanical Engineering and Research Institute for Intelligent Wearable Systems, The Hong Kong Polytechnic University, Kowloon, Hong Kong}
\affil[6]{Department of Computer Science, Hong Kong Baptist University, Kowloon Tong, Kowloon, Hong Kong}
\affil[7]{Department of Computer Science, Vanderbilt University, Nashville, TN, USA}
\affil[8]{Department of Biostatistics, Vanderbilt University Medical Center, Nashville, TN, USA}

\affil[$\dag$]{These authors contributed equally to this work.}
\affil[*]{Corresponding author.}

\begin{abstract}
Ensuring trustworthiness is fundamental to the development of artificial intelligence (AI) that is considered societally responsible, particularly in cancer diagnostics, where a misdiagnosis can have dire  consequences. Current digital pathology AI models lack systematic solutions to address trustworthiness concerns arising from model limitations and data discrepancies between model deployment and development environments. To address this issue, we developed TRUECAM, a framework designed to ensure both data and model trustworthiness in non-small cell lung cancer subtyping with whole-slide images. TRUECAM integrates 1) a spectral-normalized neural Gaussian process for identifying out-of-scope inputs and 2) an ambiguity-guided elimination of tiles to filter out highly ambiguous regions, addressing data trustworthiness, as well as 3) conformal prediction to ensure controlled error rates. We systematically evaluated the framework across multiple large-scale cancer datasets, leveraging both task-specific and foundation models, illustrate that an AI model wrapped with TRUECAM significantly outperforms models that lack such guidance, in terms of classification accuracy, robustness, interpretability, and data efficiency, while also achieving improvements in fairness. These findings highlight TRUECAM as a versatile wrapper framework for digital pathology AI models with diverse architectural designs, promoting their responsible and effective applications in real-world settings. 

\end{abstract}

\begin{document}

\maketitle

\section*{Main}
In the field of digital pathology, artificial intelligence (AI) has shown significant potential for enhancing decision support across a wide spectrum of clinical tasks, from diagnosing and distinguishing between various types of cancers to detecting subtle histopathological changes that may predict disease progression~\cite{bera2019artificial, rajpurkar2022ai,carrillo2024generation}. 
However, the reliability of medical AI models can be significantly compromised by their inherent limitations, such as weak ability in measuring uncertainty and controlling error rates. These issues are further compounded by data discrepancies between model development and deployment environments, including variations in patient demographics, disease characteristics, data acquisition techniques, staining protocols, and clinical practices. Addressing these challenges is essential to ensure model reliability, uphold patient safety, and establish trustworthiness in AI systems for medical applications~\cite{thirunavukarasu2023large,dvijotham2023enhancing}.

Rather than simply training AI models to produce the most likely outcome for patients as a point estimate---a practice that often lacks context and can lead to inaccurate interpretations---researchers have sought to quantify the confidence associated with AI generated results. This additional layer of insight can enable users to better interpret the reliability of a model's prediction or classification, helping inform when to accept the results~\citep{begoli2019need, chua2023tackling, banerji2023clinical}. Multiple uncertainty quantification (UQ) approaches have been recently integrated into AI development to characterize confidence~\citep{gal2016dropout,abdar2022need,kompa2021second,luo2023calibrated}.
For instance, Olsson and colleagues~\cite{olsson2022estimating} developed a conformalized deep convolutional neural network ensemble to generate all likely cancer diagnoses in biopsies and demonstrated the value of uncertainty estimates in flagging classifications with insufficient confidence. Additionally, Dolezal and colleagues~\cite{dolezal2022uncertainty} applied a Monte Carlo Dropout technique to quantify the uncertainty in classifications distinguishing between lung adenocarcinoma and squamous cell carcinoma, facilitating the recognition of high- versus low-confidence outputs to address domain shifts. And, most recently, Sun and colleagues~\cite{sun2024tissue} introduced TISSUE, a general framework for estimating uncertainty in spatial gene expression predictions that works by creating well-calibrated prediction intervals.

Despite advances in quantifying uncertainty, the current collection of approaches do not adequately align with several of the essential properties needed for safeguarding practical utilization of pathology AI, and medical AI more broadly~\cite{tran2022plex}. First, an ideal UQ approach should consistently characterize a model's confidence across varying levels of data risk. 
Beyond capturing uncertainty for in-domain (In-D) data, it should reliably identify out-of-domain (OOD) inputs, address potential distribution shifts between data used during model development and data encountered after deployment, and recognize regions within a slide that could adversely impact inference. At the present time, AI systems generally do not systematically address these challenges.
Second, the claimed confidence interval (e.g., a set of likely cancer subtypes) or the error rate established for real-world inference must closely align with the observed outcomes. For example, if a model predicts with 95\% confidence, then the true value should fall within the predicted range in 95\% of the cases. Although existing methods, such as Platt~\citep{niculescu2005predicting,kuleshov2018accurate,palmer2022calibration} and temperature scaling~\citep{guo2017calibration,ding2021local}, seek to improve calibration, they neither ensure consistency between expected and observed confidence nor objectively represent the model's confidence. 
Notably, these methods lack the ability to abstain from inputs that are difficult to infer accurately, a practice widely advocated for reliable medical AI with human in the loop~\cite{dvijotham2023enhancing}. Third, when estimating uncertainty, an effective UQ approach should enhance interpretability without adding substantial computational overhead beyond its deterministic counterpart (i.e., the original model without UQ). This is essential for medical imaging applications, where strict time constraints and resource limitations demand rapid and efficient processing without compromising reliability. Most UQ approaches rely on ensemble-based estimations, which require a large number of repeated inferences to be made using the same pathology AI models to estimate uncertainty for each patch (or tile) in a single slide. Given that these models typically contain millions of parameters, the resulting computational demand for generating uncertainty and interpretability can become prohibitively large, hindering their usability in practice~\citep{wilson2020bayesian}.

\begin{figure}[!ht]
    \centering
    \includegraphics[scale=0.30]{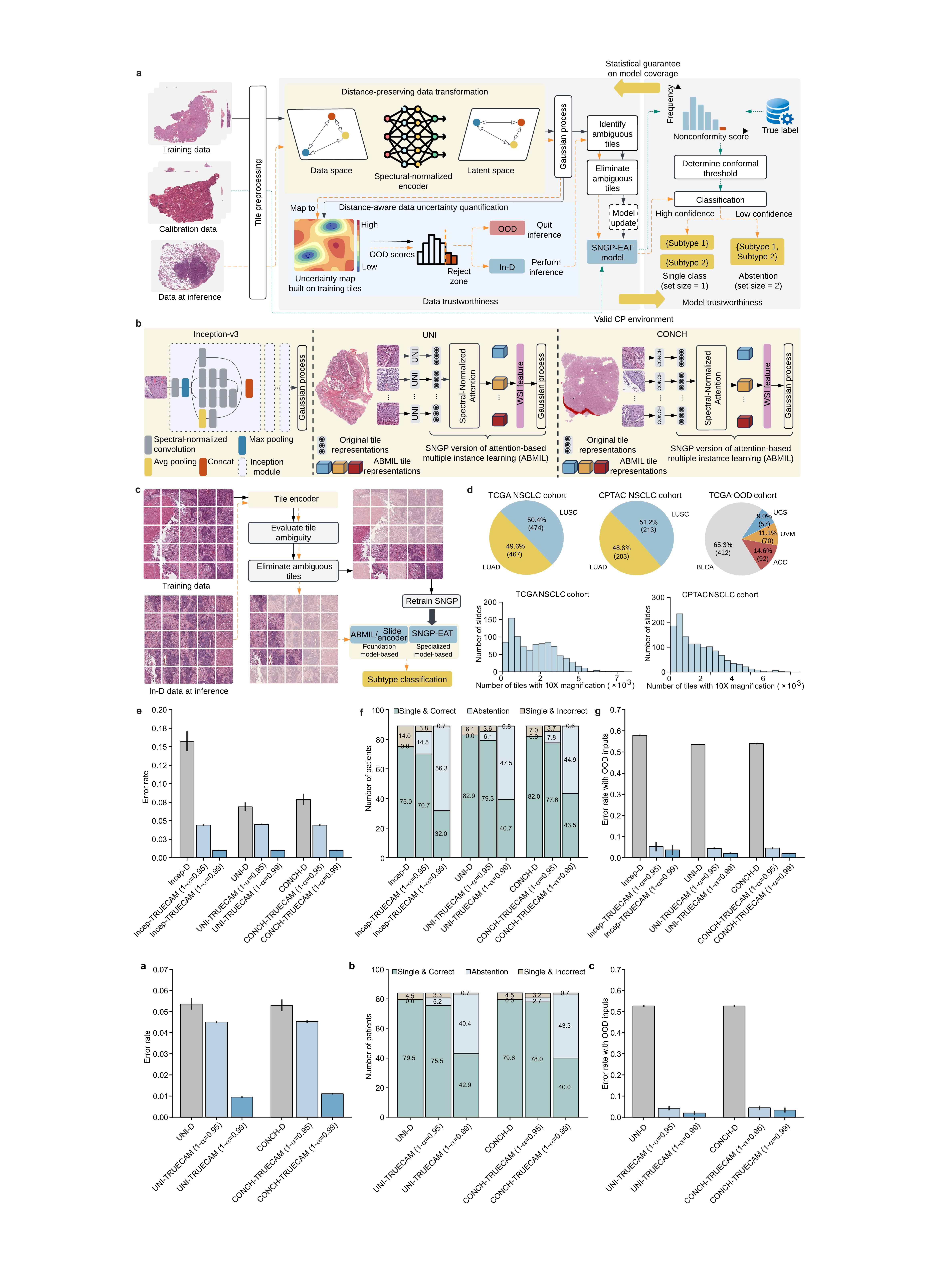}
    \captionsetup{labelformat=empty}
    \caption{}  
    \label{fig:overview}
\end{figure}

In light of these issues, we set out to equip pathology AI with a formal, principled, and scalable UQ, as well as demonstrate how it empowers trustworthy subtyping of non-small-cell lung cancers (NSCLC)~\cite{travis2011international} using whole-slide images (WSI). We decompose the trustworthiness of medical AI into two constituent parts: 1) data trustworthiness, which ensures that the input data during deployment aligns with the model’s training scope and allows ambiguous slide regions to be excluded, and 2) model trustworthiness, which offers valid confidence intervals with a customizable level of coverage, ensuring the true label is covered in a specified proportion of classifications. To do so, we developed TRUECAM (Fig.~\ref{fig:overview}a), a framework that provides \textbf{TR}ustworthiness-focused, \textbf{U}ncertainty-aware, \textbf{E}nd-to-end \textbf{CA}ncer diagnosis with \textbf{M}odel-agnostic capabilities. TRUECAM consists of three components designed to simultanuously ensure data and model trustworthiness: 1) a spectral-normalized neural Gaussian process (SNGP) to establish informative data representation and uncertainty quantification, 2) an elimination of ambiguous tiles (EAT) mechanism for filtering out noisy patches from slides, and 3) conformal prediction (CP) to enable a statistically guaranteed error rate. We then conducted a wide range of experiments with several large cancer datasets, utilizing a widely adopted specialized model (i.e., Inception-v3~\cite{szegedy2015going, kers2022deep}) and multiple general-purpose foundation models (including UNI~\cite{chen2024towards}, CONCH~\cite{lu2024visual}, Prov-GigaPath~\cite{xu2024whole}, and TITAN~\cite{ding2024multimodal}). Our results indicate that TRUECAM provides significant improvements in NSCLC subtyping accuracy and often enhances fairness in model performance across demographic groups without explicitly enforcing it into the model training process. 
Our analysis further show that TRUECAM is robust to OOD input and distribution shift, delivers more informative confidence intervals with statistical guarantee, enhances interpretability, and substantially reduces the computational inference burden. Together, these observations position TRUECAM as a general framework that can be integrated with medical AI systems of various sizes, architectures, purposes, and complexities to support their responsible and trustworthy application.

\addtocounter{figure}{-1}
\begin{figure} [t!]
  \caption{Overview of TRUECAM. TRUECAM is a versatile, model-agnostic digital pathology AI framework for reliable non-small-cell lung cancer (NSCLC) subtyping, achieving trustworthiness by substantially reducing errors, detecting OOD data and controlling distribution shifts pre-inference, identifying and eliminating ambiguous slide regions, and ensuring true label coverage with statistical guarantees via abstention on uncertain inputs. \textbf{a}, The architecture of TRUECAM designed to ensure both data and model trustworthiness. \textbf{b}, Customization illustration of TRUECAM for deep learning models of varying architecture, complexity, and purpose, including Inception-v3, UNI, and CONCH. \textbf{c}, Illustration of eliminating ambiguous tiles in TRUECAM for slide inference. \textbf{d}, Overview of the TCGA and CPTAC NSCLC datasets, as well as the out-of-domain dataset built from other cancer types within TCGA. \textbf{e}, TRUECAM significantly reduced NSCLC subtyping error rates across all model types (denoted using suffix ``-TRUECAM'') compared to their original deterministic versions (denoted using suffix ``-D''), adhering to the pre-specified true label coverage  1-$\alpha$. \textbf{f}, Patient-level classification breakdown for models with and without TRUECAM. \textbf{g}, TRUECAM achieved significantly lower error rates in real-world NSCLC subtyping scenarios involving a 1:1 mix of in-domain and out-of-domain inputs. Results shown in \textbf{e}-\textbf{g} are based on the TCGA dataset. See Extended Data Fig.~\ref{fig:performance-summary-cptac} for evaluations based on the CPTAC dataset and two other foundation models (Prov-GigaPath and TITAN). All mean values and 95\% confidence intervals are based on 20 independently trained models, each with 500 conformal prediction evaluations. OOD, out-of-domain; In-D, in-domain; SNGP, spectral-normalized neural Gaussian process; EAT, elimination of ambiguous tiles; WSI, whole-slide image; LUAD, lung adenocarcinoma; LUSC, lung squamous cell carcinoma; BLCA, bladder urothelial carcinoma; USC, uterine carcinosarcoma; UVM, uveal melanoma; ACC, adrenocortical carcinoma; Incep, Inception-v3; D, Deterministic. }
\end{figure}

\section*{Results}

\subsection*{An overview of TRUECAM} 

This study focuses on distinguishing between two main types of of NSCLC--lung adenocarcinoma (LUAD) and lung squamous cell carcinoma (LUSC)---each of which is defined by distinct biological, morphological, molecular, and prognostic features that inform which therapeutic approach to undertake for a patient~\cite{travis2011international}. WSIs of lung tissue specimens were used for this specific task. In this paper, we use the term ``prediction'' interchangeably with ``classification'' where needed to align with the established terminology convention in conformal prediction.

We designed the TRUECAM framework (Fig.~\ref{fig:overview}a) to wrap around deep learning models of various architectures, sizes, and purposes for imaging tasks (Fig.~\ref{fig:overview}b). TRUECAM employs SNGP to perform distance-preserving transformations on input data and efficiently estimate data uncertainty, enhancing data trustworthiness by enabling OOD detection and data shift control before model inference. Furthermore, CP is applied on top of SNGP to calibrate model predictions (represented as a set of values, i.e., prediction set), providing a statistical guarantee on the model's coverage of true labels. These components contribute complementary functions to assure the trustworthiness of 
AI for NSCLC subtyping. Specifically, SNGP establishes a valid foundation for CP to operate. This is important because CP generally requires that data in the deployment environment comes from the same underlying distribution as those in the development environment, known as data exchangeability---an assumption that might not hold true in the real-world setting, where OOD data and distribution shift are common. Meanwhile, CP calibrates the outputs of SNGP to ensure the statistical validity of model coverage. 
Notably, the distance-preserving data transformation by SNGP facilitates an effective 
identification of tiles that only add ambiguity to NSCLC subtyping. EAT can then be applied to enhance the supervisory signal---leading to a more reliable model---while significantly reducing the computational load during inference (Fig.~\ref{fig:overview}c). 
The details of model design and implementation are described in the Methods.

In this study, we relied on two digital pathology datasets of NSCLC stained with hematoxylin and eosin (H\&E) (Fig.~\ref{fig:overview}d), both of which are large-scale, multi-institutional initiatives in the United States: 1) 941 WSIs from The Cancer Genome Atlas (TCGA), and 2) 1,306 WSIs from the Clinical Proteomics Tumor Analysis Consortium (CPTAC). Additionally, we constructed an OOD dataset from the same source as TCGA, which incorporates 631 WSIs of non-lung cancers. For simplicity, we refer to these three datasets as TCGA, CPTAC, and TCGA-OOD. The details for these datasets are provided in the Methods.

To assess the effectiveness of TRUECAM, we evaluated two types of AI models. We utilized Inception-v3, a widely used convolutional neural network architecture for image inference~\cite{szegedy2015going, kers2022deep}, as a representative classifier for specialized medical imaging tasks within the TRUECAM framework. In this context, we used TCGA for model training and internal validation, while we used CPTAC as an independent external validation dataset to assess TRUECAM's effectiveness in scenarios requiring model transferability. We further investigated
four digital pathology foundation models that exemplify the recent advancements in large, general-purpose models in the field: 1) UNI~\cite{chen2024towards}, 2) CONCH~\cite{lu2024visual}, 3) Prov-GigaPath~\cite{xu2024whole}, and 4) TITAN~\cite{ding2024multimodal}. Since foundation models are pretrained on diverse datasets, offering strong knowledge transferability as an image encoder, and downstream task-specific classifiers are typically lightweight and easy to train, our evaluation focused solely on the setting where each healthcare organization trains and evaluates its own site-specific model using its data. Accordingly, TCGA and CPTAC were utilized to mimic this setting, such that we assumed each dataset represents a distinct healthcare organization, independently supporting the required model training and validation. Our evaluation primarily focused on UNI and CONCH, with additional results provided for Prov-GigaPath and TITAN to showcase the generalizability of TRUECAM's advantages.

Our evaluation yielded several notable findings. First, TRUECAM significantly reduced NSCLC subtyping error rates (defined as the fraction of incorrect patient-level classifications) across all three models—Inception-v3, UNI, and CONCH—compared to their original versions, denoted as Deterministic (or D) (Fig. \ref{fig:overview}e). Defining $\alpha$ as the desired maximum error level (or significance level), applying TRUECAM to Inception-v3 (referred to as Incep-TRUECAM) with coverage of 1-$\alpha$=0.95 and 1-$\alpha$=0.99 resulted in error rate reductions by 72.0\% and 93.8\%, respectively. Notably, these empirical error rates  closely complemented the desired coverage levels that reflect adaptable error tolerance for individual NSCLC subtyping services. Second, TRUECAM demonstrated an ability to identify and abstain from reasoning about challenging inputs, allowing such inputs to be deferred to pathologists.
As shown in Fig. \ref{fig:overview}f, with coverage set at 1-$\alpha$=0.95, Incep-TRUECAM abstained on approximately 15 patients, reducing the number of misclassified patients from 14 to fewer than 4. Third, the foundation model-based classifiers exhibited more promising results than Inception-v3, with inherently lower error rates and fewer abstentions
to achieve the same desired coverage (Fig. \ref{fig:overview}e,f). 
Fourth, in simulated real-world settings where OOD slides may be unknowingly submitted for NSCLC subtyping, directly deploying the original versions of the considered models 
often incorrectly classified non-lung cancer slides as LUAD or LUSC. By contrast, TRUECAM reliably detected these OOD inputs prior to model inference, while maintaining the same 
error rates
as the scenarios involving only In-D inputs (Fig. \ref{fig:overview}g). We observed similar findings in the evaluation of Prov-GigaPath and TITAN (Extended Data Fig.~\ref{fig:performance-summary-cptac}). 

The following sections report on a systematic evaluation of TRUECAM's performance in various dimensions and the effectiveness of its core modules, using Inception-v3 as the primary example. We then extend the analysis to foundation models UNI and CONCH.

\subsection*{Integration of distance-aware uncertainty estimation improves NSCLC subtyping performance}

We trained three deep neural network models, all based on the Inception-v3 architecture, to discriminate between LUSC and LUAD: 1) a deterministic model without UQ, referred to as Deterministic, which represents the original Inception-v3, 2) a Monte Carlo Dropout-based model (referred to as MC Dropout) to enable UQ~\cite{dolezal2022uncertainty}, and 3) a model that utilizes SNGP to quantify uncertainty (referred to as SNGP). We provide model details in the Methods. Following the design in ~\cite{dolezal2022uncertainty}, all models were trained at the tile level, with slide-level classifications obtained by averaging tile-level outputs across each WSI. For patients with multiple WSIs, the final patient-level diagnosis was derived by averaging the slide-level classifications across all WSIs for that patient.

\begin{figure}[!ht]
    \centering
    \includegraphics[scale=0.28]{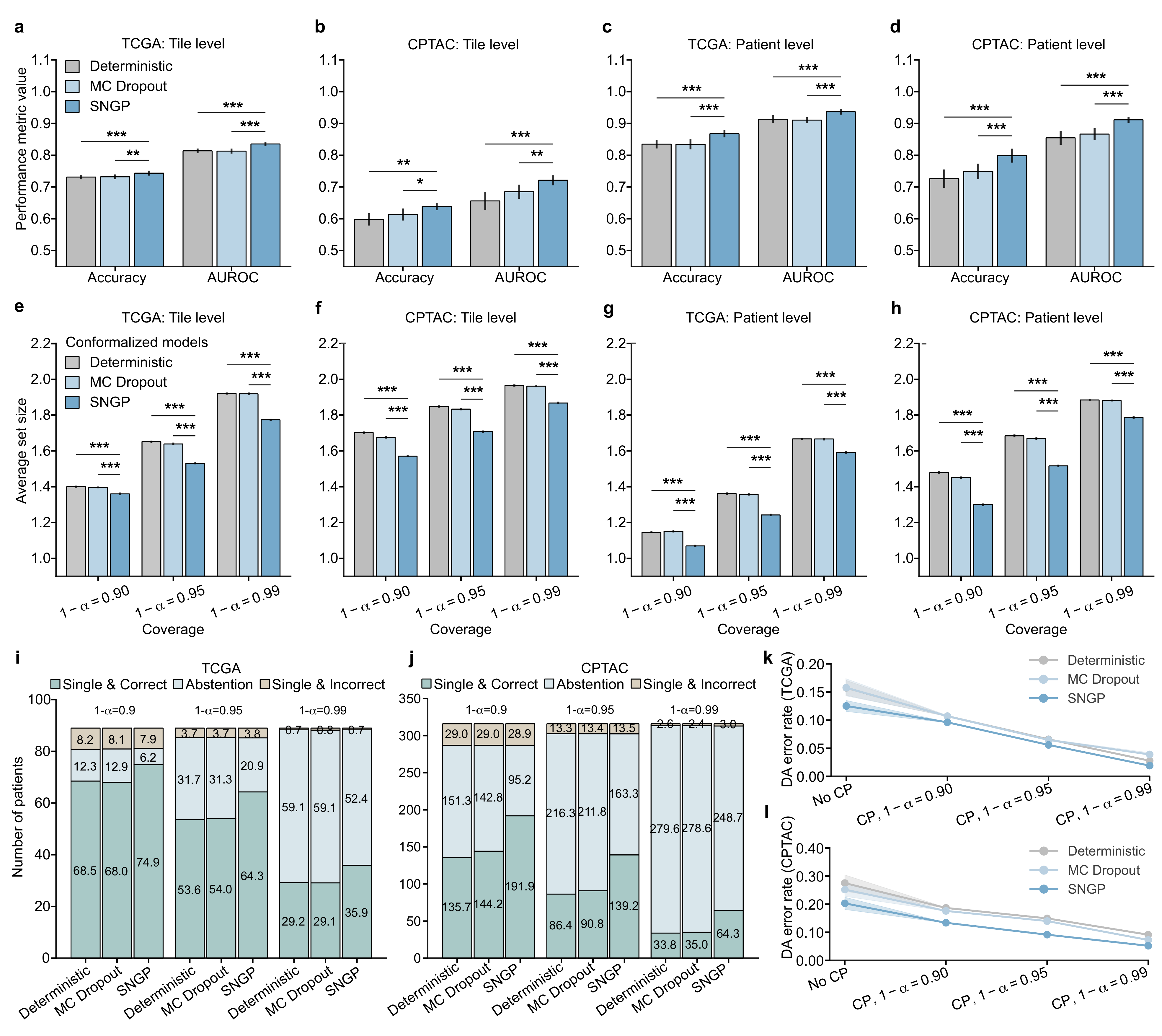}
    \caption{NSCLC subtyping performance of three Inception-v3-based deep neural network models and their conformalized counterparts. \textbf{a}, Tile-level performance with respect to the TCGA dataset in terms of classification accuracy and area under receiver operator curve (AUROC). \textbf{b}, Tile-level performance evaluated on the CPTAC dataset. \textbf{c}, Patient-level performance evaluated on the TCGA dataset. \textbf{d}, Patient-level performance evaluated on the CPTAC dataset. \textbf{e}, Tile-level prediction set size on TCGA for three distinct values of error level $\alpha$. \textbf{f}, Tile-level prediction set size on CPTAC. \textbf{g}, Patient-level prediction set size on TCGA. \textbf{h}, Patient-level prediction set size on CPTAC. \textbf{i}, Patient-level classification breakdown for three conformalized models on the TCGA testing dataset. \textbf{j}, Patient-level classification breakdown on CPTAC. \textbf{k}, Patient-level DA error rate on TCGA before and after activating CP. \textbf{l}, Patient-level DA error rate on CPTAC before and after activating CP. Mean values and the corresponding 95\% confidence intervals in \textbf{a}-\textbf{d} are derived from 20 independently trained models, evaluated using a combination of calibration and testing data. \textbf{e}-\textbf{l} are based on 20 independently trained models, each with 500 CP evaluations (See details in the Methods). One-sided Wilcoxon signed-rank test is utilized to calculate \textit{p} values. *** $p<0.001$; ** $p<0.01$; * $p<0.05$. SNGP, spectral-normalized neural Gaussian process; CP, conformal prediction; DA, definitive-answer.}
    \label{fig:cp_comparison}
\end{figure}

We deactivated TRUECAM's CP functionality and EAT to assess the effectiveness of SNGP. At the tile level (using tiles as subtyping units), SNGP consistently outperformed Deterministic on the TCGA dataset in both accuracy ($p < 0.001$) and the area under receiver operator curves (AUROC) ($p < 0.001$)  (Fig.~\ref{fig:cp_comparison}a). In addition, when evaluated on TCGA, SNGP significantly outperformed MC Dropout in terms of both accuracy ($p < 0.01$) and AUROC ($p < 0.001$). When evaluated on the external CPTAC dataset, SNGP exhibited similar advantages over Deterministic and MC Dropout (Fig.~\ref{fig:cp_comparison}b). At the patient level (using patients as subtyping units), the performance improvement enabled by SNGP was more pronounced. In this situation, Deterministic achieved an average accuracy of 0.843 and AUROC of 0.926 for TCGA, whereas SNGP achieved an accuracy of 0.875 and AUROC of 0.950 (Fig.~\ref{fig:cp_comparison}c). Moreover, integrating SNGP into Inception-v3 led to a significant performance advantage over MC Dropout (accuracy: $p < 0.001$; AUROC: $p < 0.001$). Further, the performance gain enabled by SNGP generalized well to the external CPTAC dataset. Specifically, SNGP outperformed MC Dropout by 4.9\% ($p < 0.001$) and Deterministic by 7.2\% ($p < 0.001$) in accuracy, and exceeded MC Dropout by 4.5\% ($p < 0.001$) and Deterministic by 5.7\% ($p < 0.001$) in AUROC (Fig.~\ref{fig:cp_comparison}d).

\subsection*{Conformalized SNGP establishes statistical coverage guarantee while enhancing NSCLC subtyping efficiency} 
We enabled CP in TRUECAM and then assessed its effectiveness in calibrating SNGP-based NSCLC subtyping by generating a prediction set for each patient. A set size of 2 indicates that the model lacks sufficient confidence to definitively classify input data as either LUAD or LUSC, effectively resulting in abstention. Two broadly used measures in the literature, validity and efficiency ~\cite{shafer2008tutorial,balasubramanian2014conformal,romano2019conformalized} were evaluated. 
We consider a model to be valid at an error level $\alpha$ when the proportion of the prediction set containing the true subtype (i.e., model coverage) is at least $1-\alpha$. Efficiency is determined by the size of the prediction sets---we deem a model to be more efficient when it produces smaller prediction sets, provided that validity is maintained.  This is because more concise and informative outputs are generally preferred by both service providers and patients. 

There are several findings we wish to highlight. First, all three of the models achieved the validity requirements at both the tile and patient levels across the TCGA and CPTAC testing datasets for three distinct error levels (Supplementary Tables~\ref{tab:ec-tile-tcga-cptac},~\ref{tab:ec-patient-tcga-cptac}). 
Second, for each considered coverage $1-\alpha$, conformalized SNGP consistently produced significantly smaller prediction sets than conformalized MC Dropout and Deterministic at both the tile and patient levels (Fig. \ref{fig:cp_comparison}e,g). 
Such an advantage generalized well to the external CPTAC dataset at both the tile and patient levels (Fig. \ref{fig:cp_comparison}f,h). More concretely, at $\alpha=0.05$, conformalized SNGP produced 41,570 more tiles 
with a single NSCLC subtype output (set size of one) on average, which was a 31.4\% increase over conformalized MC Dropout. In other words, when validity was ensured, conformalized SNGP exhibited a significantly higher efficiency in terms of the number of informative outputs in NSCLC subtyping than other models (Supplementary Tables~\ref{tab:tcpp-tcga-cptac},\ref{tab:pcpp-tcga-cptac}). Third, as expected, lowering the value of $\alpha$, which decreases the tolerance for model miscoverage, resulted in larger prediction sets across all models, datasets, and classification levels (Fig. \ref{fig:cp_comparison}e-h). In other words, more tiles and patients were classified as ``unsure'' (i.e., abstention) to reduce the likelihood of classifications containing only a single, but incorrect, subtype.

We further investigated the models' patient-level classifications, which include the following categories 1) single and correct, 2) single but incorrect, and 3) uncertain between subtypes (abstention). For all considered error levels, conformalized MC Dropout achieved similar patient-level performance as Deterministic in terms of patient numbers under each category (Fig.~\ref{fig:cp_comparison}i,j). By contrast, conformalized SNGP consistently produced more single and correct classifications in both of the TCGA and CPTAC testing datasets than the other two models, while incurring a similar number of single but incorrect classifications (Fig.~\ref{fig:cp_comparison}i,j). Specifically, at $\alpha=0.01$, conformalized SNGP classified approximately six more patients with a single and correct NSCLC subtype than conformalized MC Dropout, reflecting a 23.4\% increase (Fig. \ref{fig:cp_comparison}i). This advantage was further amplified in the CPTAC testing dataset, where it classified around 30 more patients with single and correct subtype at $\alpha=0.01$, corresponding to a 83.7\% increase than conformalized MC Dropout (Fig.~\ref{fig:cp_comparison}j). As an artifact of bounded error rates, the increase in correctly identified patients by conformalized SNGP was accompanied by a comparable reduction in classifications with abstention. These findings collectively highlight enhanced overall performance and greater efficiency of TRUECAM, as evidenced by the increased ratio of single and correct classifications, underscoring its superior practical utility.

Next, we examined the capability of TRUECAM to manage errors when committing to a single subtype prediction. 
We define the \textit{definitive-answer error rate} (referred to as DA error rate) as the proportion of patients with single but incorrect subtype designation among all patients with single subtype classification.  Before activating CP, TRUECAM demonstrated the lowest error rate for both TCGA and CPTAC (Fig.~\ref{fig:cp_comparison}k,l). In other words, when the model outputs were not conformalized (i.e., classifications were binary, based solely on the highest probability between subtypes), SNGP outperformed both MC Dropout and Deterministic, a finding consistent with the results shown in Fig.~\ref{fig:cp_comparison}c,d. When CP allows a model to abstain on uncertain inputs, all models achieved a reduction in DA error rate, with a steady decline as the significance level $\alpha$ decreases. Notably, at $\alpha=0.01$, CP induced a DA error rate of 1.9\% in TCGA, representing an 84.8\% reduction compared to using SNGP alone. In other words, SNGP, without CP, made approximately one error for every eight patients, whereas with CP at $\alpha=0.01$, SNGP made only one error for every 100 patients on average. These findings, along with similar trends observed across other models, demonstrated that CP provided a significant advantage by universally reducing the DA error rate across various models. Additionally, for all considered coverage values $1-\alpha$, conformalized SNGP exhibited the lowest DA error rates, indicating that TRUECAM delivered more accurate classifications than other models.

\subsection*{Eliminating ambiguous tiles (EAT) concurrently augments classification and CP performance in NSCLC subtyping} 

In clinical inference using a WSI, not all regions provide diagnostic value to pathologists (e.g., areas with normal, non-cancerous cells). However, the current practice in digital pathology typically involves utilizing all tiles—often numbering in the hundreds to tens of thousands from a single WSI (Fig.~\ref{fig:overview}d)—paired with coarse-grained slide-level labels for training diagnostic models~\cite{coudray2018classification,campanella2019clinical,lu2021data,chen2021annotation,claudio2024mapping}. Such a weakly supervised learning process inherently introduces noise, as many normal tiles may be inaccurately associated with tumor labels~\cite{jiang2024transformer}. This practice can lead to reliability issues and inefficiencies as models are forced to learn from irrelevant or non-informative regions (even with attention mechanisms in place). These, in turn,  dilute the signal necessary for accurate and efficient NSCLC subtyping, which undermines data trustworthiness. 

\begin{figure}[!ht]
    \centering
    \includegraphics[scale=0.23]{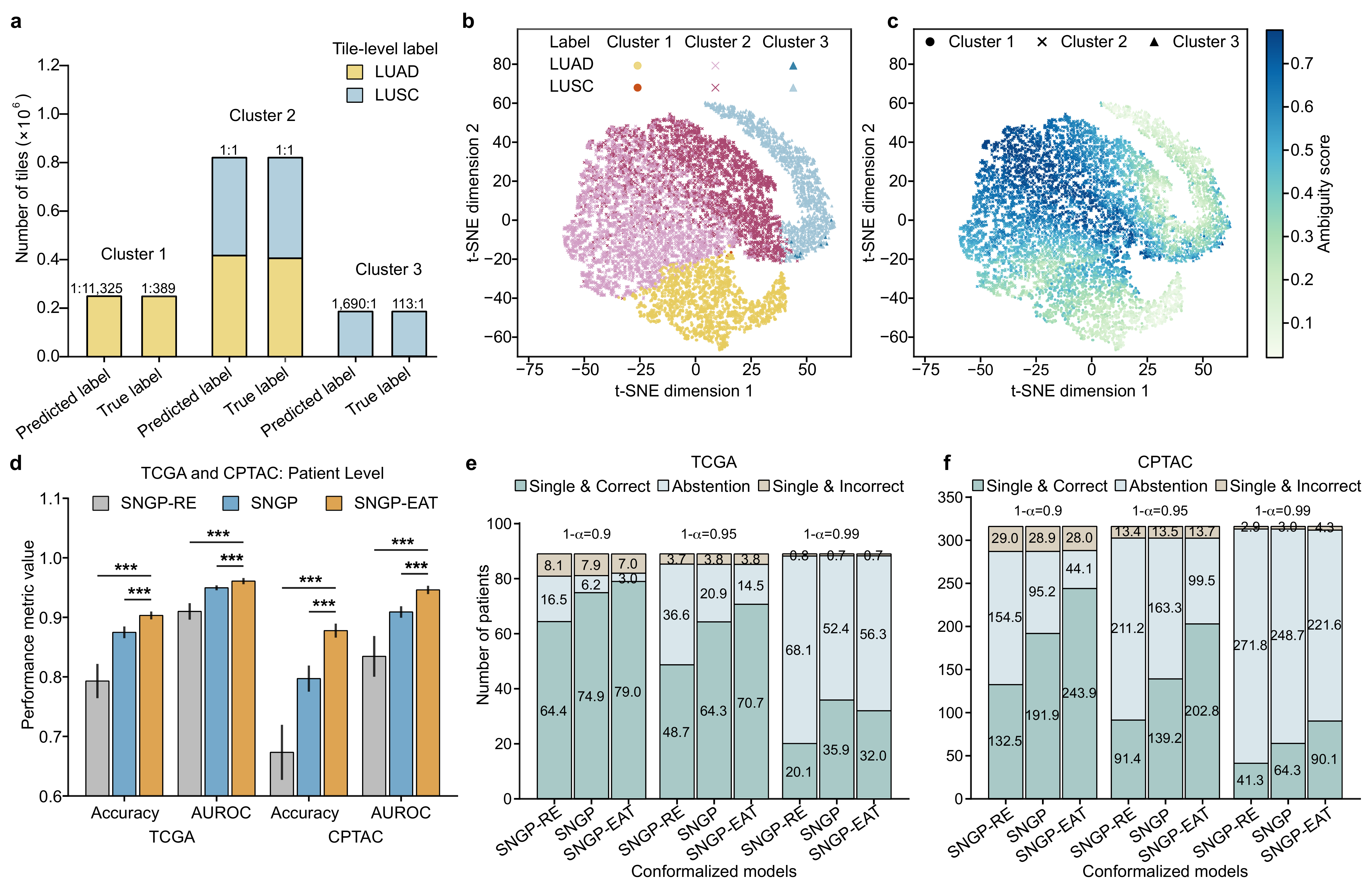}
    \caption{Evaluation of eliminating ambiguous tiles (EAT). \textbf{a}, The count and ratio of tiles predicted by SNGP as LUSC versus LUAD within each tile cluster identified using $k$-means on SNGP-based tile representations, along with the corresponding true labels derived from slide-level diagnosis. \textbf{b}, T-distributed stochastic neighbor embedding (t-SNE)-based illustration of tile clusters and true labels. \textbf{c}, Tile-wise ambiguity scores on top of the t-SNE visualization. For the sake of illustration, we  randomly sampled 10,000 tiles to visualize the t-SNE distributions in \textbf{b} and \textbf{c}. \textbf{d}, Patient-level classification performance on the TCGA and CPTAC testing datasets. \textbf{e}, Patient-level classification breakdown of three conformalized SNGP-based models on TCGA. \textbf{f}, Patient-level classification breakdown on CPTAC. Mean values and the corresponding 95\% confidence intervals in \textbf{d}-\textbf{f} are based on 20 independently trained models, each with 500 CP evaluations (See details in the Methods). One-sided Wilcoxon signed-rank test is utilized to calculate \textit{p} values. *** $p<0.001$. LUAD, lung adenocarcinoma; LUSC, lung squamous cell carcinoma; SNGP, spectral-normalized neural Gaussian process; RE, random elimination; EAT, elimination of ambiguous tiles.}
    \label{fig:EAT_summary}
\end{figure}

To filter out non-informative tiles for the NSCLC subtyping task, we applied $k$-means clustering to the tile representations of the TCGA training data extracted by SNGP. We calculated the Silhouette coefficient~\cite{shahapure2020cluster} and determined $k=3$ as the optimal number of clusters. We observed that one cluster contained mostly tiles from WSIs with the true label of LUAD,  another cluster contained mostly tiles with the true label of LUSC, and the final cluster was not dominated by either subtype (Fig.~\ref{fig:EAT_summary}a). Using t-distributed stochastic neighbor embedding (t-SNE) visualization (Fig.~\ref{fig:EAT_summary}b), we 
observed that the clustering pattern consistently aligned with the label distribution depicted in Fig.~\ref{fig:EAT_summary}a.

Subsequently, we measured the ambiguity of all of the tiles to quantify model discernibility (formally defined in the Methods) and observed that the ambiguity scores were significantly higher in the cluster lacking clear label dominance (referred to as ambiguous cluster) than the other two clusters (Fig.~\ref{fig:EAT_summary}c). This suggests that the tiles in the ambiguous cluster may only add ambiguity to NSCLC subtyping when aggregating tile-level classifications. Motivated by these observations, we hypothesized that eliminating ambiguous tiles through the EAT process would strengthen the supervisory signals for model training and lead to more accurate subtyping of NSCLC. To test this hypothesis, we compared the classification and CP performance of TRUECAM before and after activating EAT. To do so, we removed the training tiles from the ambiguous cluster (accounting for 66.7\% of all training tiles) and used the remainder of the data to train a new classification model (Fig.~\ref{fig:overview}c), which we refer to as SNGP-EAT. For new WSI inference, tiles from a new slide with their latent representations falling within the ambiguous cluster were excluded from the NSCLC subtyping task using SNGP-EAT.

In addition to comparing SNGP-EAT with SNGP, we established another baseline model, referred to as SNGP-RE, which follows the same procedure as SNGP-EAT but eliminates an equal number of tiles that are selected uniformly at random. SNGP-EAT achieved a 2.83\% improvement in patient-level classification accuracy compared to SNGP ($p < 0.001$) on TCGA and an 8.05\% increase ($p < 0.001$) on CPTAC (Fig.~\ref{fig:EAT_summary}d, Supplementary Table~\ref{tab:iacp}). By contrast, SNGP-RE consistently performed worse than both SNGP and SNGP-EAT in terms of accuracy and AUROC (Extended Data Fig.~\ref{fig:extended_data_figure1}a-d). Moreover, EAT improved CP efficiency by producing significantly smaller prediction set sizes (Extended Data Fig.~\ref{fig:extended_data_figure1}e-h, Supplementary Tables~\ref{tab:tcpp-tcga-cptac},~\ref{tab:pcpp-tcga-cptac}) and generated more single and correct classifications in both the TCGA and CPTAC cohorts (Fig.~\ref{fig:EAT_summary}e,f), except for the TCGA setting with $\alpha=0.01$, while yielding a comparable number of single but incorrect classifications as the models without EAT. Meanwhile, a comparison of DA error rates also favored SNGP-EAT (Extended Data Fig.~\ref{fig:extended_data_figure1}i,j, Supplementary Tables~\ref{tab:error-rate-cpp-tcga-cptac},\ref{tab:pl-error-rate-tcga}).

\subsection*{TRUECAM achieves fairer NSCLC diagnosis compared to other methods}

We investigated how TRUECAM affects fairness by comparing it to a set of baseline models. We first evaluated the classification performance gap, defined as the maximum difference in accuracy among patient subgroups based on race and sex (Supplementary Table \ref{tab:data_summary}). This was assessed in scenarios without CP. We then extended this evaluation to fairness in terms of average set sizes of CP across these subgroups. A large difference in set sizes between two subgroups suggests greater uncertainty in the model's predictions for one subgroup compared to the other. 

\begin{figure}[!ht]
    \centering
    \includegraphics[scale=0.215]{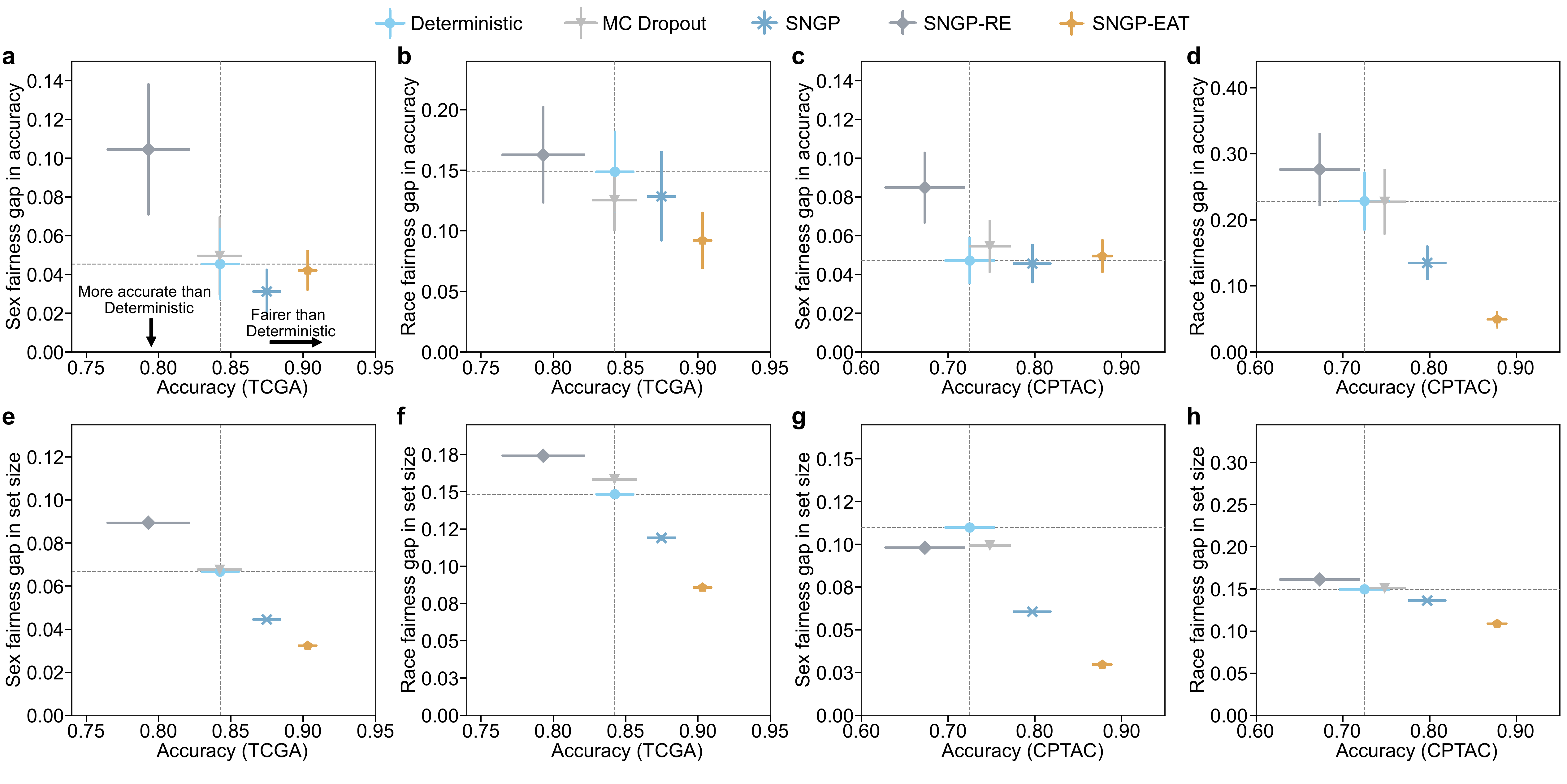}
    \caption{Fairness evaluation of TRUECAM based on metric value differences between the best- and worst-performing demographic subgroups against overall accuracy of NSCLC subtyping. \textbf{a-d}, Sex- and race-wise accuracy gap versus overall accuracy for discriminating between LUAD and LUSC on TCGA (n=189) and CPTAC (n=416) with CP deactivated. \textbf{e-h}, At $\alpha = 0.1$, sex- and race-wise set size gap versus overall accuracy on TCGA and CPTAC with CP activated. Evaluations on Deterministic or its conformalized version are marked as an overall baseline. Additional results are provided in Supplementary Tables~\ref{tab:gender_accuracy_gap_TCGA}-\ref{supptable:gender_CP_CPTAC}.}
    \label{fig:classification_performance_gap}
\end{figure}

 SNGP and SNGP-EAT consistently exhibited higher overall classification accuracy than Deterministic and MC Dropout, while also improving fairness in accuracy across racial and sexual subgroups when Deterministic's fairness gap was relatively high or preserving fairness when the gap was relatively low  (Fig.~\ref{fig:classification_performance_gap}a-d). Notably, SNGP-EAT achieved the highest overall accuracy on both of the internal and external datasets and exhibited the smallest fairness gaps in accuracy across racial subgroups. Specifically, SNGP-EAT led to a 38.1\% and 78.3\% reduction in fairness gaps across racial groups in TCGA and CPTAC, respectively, compared to Deterministic (Supplementary Tables~\ref{tab:race_accuracy_gap_TCGA}-\ref{tab:race_accuracy_gap_CPTAC}). By contrast, randomly removing tiles through SNGP-RE did not result in a significant improvement in overall accuracy and fairness, again highlighting the benefit of EAT.




The advantages of conformalized SNGP and SNGP-EAT were more pronounced for fairness in the average set sizes of CP (Fig.~\ref{fig:classification_performance_gap}e-h). Remarkably, conformalized SNGP-EAT (i.e., TRUECAM), consistently outperformed all other models by a large margin in terms of the average set size gap across sexual and racial subgroups. For example, at $1-\alpha=0.90$, compared to conformalized Deterministic, conformalized SNGP-EAT achieved a reduction of 42.2\% in the set size gap across racial subgroups on TCGA and reduces such gap by 27.4\% on CPTAC (Figs.~\ref{fig:classification_performance_gap}f,h, Supplementary Tables~\ref{tab:race_CP_gap_TCGA}-\ref{tab:race_CP_gap_CPTAC}). 
With respect to sex, conformalized SNGP-EAT reduced the fairness gap for TCGA and CPTAC by 51.6\% and 73.0\%, respectively, approaching a near-perfect value (Fig.~\ref{fig:classification_performance_gap}e,g, Supplementary Tables~\ref{supptable:gender_CP_TCGA}-\ref{supptable:gender_CP_CPTAC}). These improvements were accompanied by significant gains in the overall model accuracy.


\subsection*{TRUECAM enables effective OOD detection and distribution shift control}

In pathology AI deployment, WSIs that differ from the model's training data in aspects such as cohorts, tissue types, devices, and other factors are prevalent and difficult to detect. For models equipped with CP, the claimed model coverage may no longer hold valid under such condition. This potentially undermines reliability and poses a serious risk to patient health. We now show that TRUECAM is able to safeguard CP by detecting data that is outside of the model's scope (i.e., OOD data).   

\begin{figure}[!ht]
    \raggedright
    \includegraphics[scale=0.21]{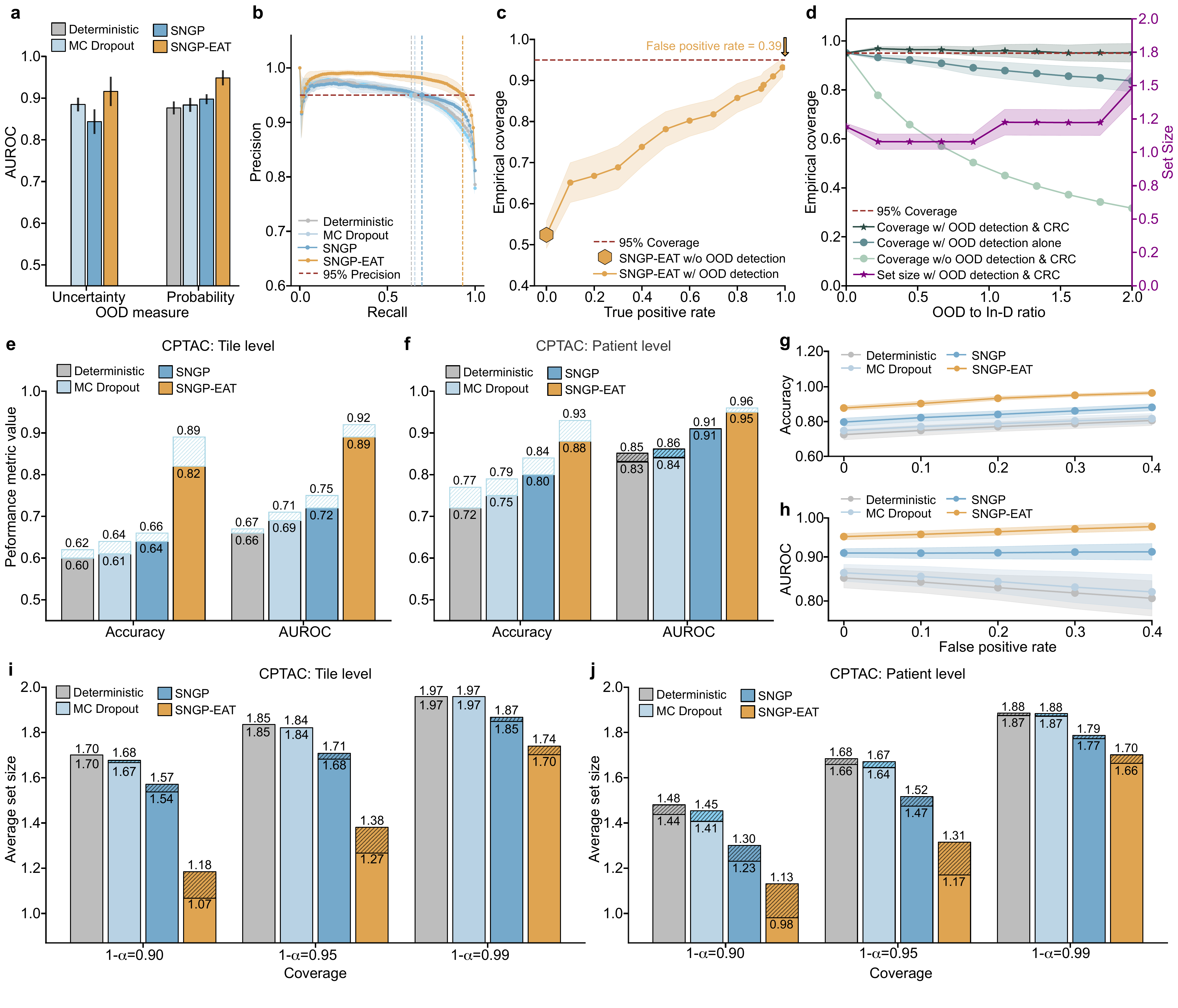}
    \caption{Evaluation of OOD detection enabled by TRUECAM. \textbf{a}, Performance of two distinct OOD scores in identifying slides outside a model's scope, using the top $\delta$=200 tiles in a slide with the lowest tile-level values for patient-level OOD score calculation. The bar for Deterministic is not shown as it does not measure uncertainty. See Supplementary Table~\ref{tab:lambda_OOD} for sensitivity analyses for varying $\delta$ values. \textbf{b}, Precision-recall curves for OOD detection using the probability-based OOD score. \textbf{c}, Empirical model coverage with $\alpha=0.05$ as a function of the true positive rate for OOD detection. \textbf{d}, Empirical coverage and prediction set size of CP for TRUECAM, compared to the scenario where conformal risk control and OOD detection were deactivated, shown as a function of the ratio of OOD to In-D data.  \textbf{e}, Impact assessment of distribution shift control on the tile-level classification performance. Striped boxes in blue and black indicate performance increase and decrease, respectively, compared to the scenario without distribution shift control. \textbf{f}, Impact assessment of distribution shift control on the patient-level classification performance. \textbf{g,h}, Accuracy and area under the receiver operating characteristic (AUROC) for evaluation on CPTAC as a function of false positive rate in OOD detection. \textbf{i}, Impact assessment of distribution shift control on the tile-level average set size. Striped boxes in black indicate improvement in set size, compared to the scenario without distribution shift control. \textbf{j}, Impact assessment of distribution shift control on the patient-level average set size. SNGP, spectral-normalized neural Gaussian process; EAT, elimination of ambiguous tiles; CRC, conformal risk control.}
    \label{fig:SNGP_CP_complementary_summary}
\end{figure}

We built the TCGA-OOD dataset that contains 698 WSIs from 631 patients with non-lung cancers (See Supplementary Table~\ref{tab:data_summary} and Methods for details), and then combined it with an In-D dataset, which integrates the TCGA calibration and testing datasets. We defined two distinct measures to quantify the degree to which an input WSI is out of the model's scope: 1) an uncertainty-based OOD score, which aggregates tile-level uncertainty as determined by the model's UQ module (e.g., Gaussian layer for SNGP and SNGP-EAT, stochastic dropout layer for MC Dropout), and 2) a probability-based OOD score, which aggregates tile-level classification probabilities, calculated as $1 - \frac{1}{N} \sum_{i=1}^{N} \max_{k} p(\hat{y}_k|\bm{x}_i)$, where $p(\hat{y}_k|\bm{x}_i)$ denotes the probability of tile $i$ being classified as label $\hat{y}_k$. 
We observed that probability-based OOD scores from SNGP-EAT exhibited the strongest ability to distinguish between In-D and OOD WSIs, achieving the highest AUROC of 0.949, significantly outperforming Deterministic (0.876), MC Dropout (0.884), and SNGP (0.898) (Fig.~\ref{fig:SNGP_CP_complementary_summary}a). At a 0.95 precision, SNGP-EAT achieved a true positive rate (TPR, i.e., recall) of 0.929, which is 46.0\%, 41.5\%, and 33.3\% higher than Deterministic, MC Dropout, and SNGP, respectively (Fig.~\ref{fig:SNGP_CP_complementary_summary}b). This observation also generalized to uncertainty-based OOD scores. Notably, for both Inception-v3-based SNGP models, probability-based OOD scores were more effective than uncertainty-based scores for identifying OOD inputs (Supplemtentary Table~\ref{tab:OOD_detection_performance_summary}). As such, the probability-based OOD score was relied upon in the subsequent investigations related to Inception-v3-based models. 

We further examined the effect of identifying and removing OOD inputs from inference enabled by TRUECAM on the empirical coverage of a model (i.e., model validity). When we randomly sampled an equal number of OOD patients to match the size of the In-D dataset, the empirical coverage of SNGP-EAT approached the intended value of 0.95 ($\alpha$=0.05) as the true positive rate (TPR) of OOD detection neared 1 (Fig.~\ref{fig:SNGP_CP_complementary_summary}c). This effectively restored data exchangeability. By contrast, without an OOD detection mechanism, the empirical coverage remained at 0.478, which fell significantly short of the target and signaled a failure in the model's reliability.

However, even with the best-performing model and OOD score, identifying 99.0\% OOD inputs came at the cost of misclassifying 39\% of In-D data as OOD (Fig.~\ref{fig:SNGP_CP_complementary_summary}c), excluding them from model inference. 
Furthermore, while the OOD threshold was set to achieve TPR = 1 in testing, this level of coverage cannot be guaranteed in deployment, where unexpected OOD data may extend beyond the threshold, thus compromising model coverage and trustworthiness.
To address these challenges, TRUECAM pairs conformal risk control (CRC) with OOD detection, aiming to provide a more reliable model coverage even in the presence of undetected OOD data. CRC adjusts model outputs dynamically by accounting for the risks posed by OOD inputs that escape detection, thereby maintaining valid empirical coverage (see Methods for details). 

We assessed TRUECAM's effectiveness in handling OOD data by simulating a realistic environment where the ratio of OOD to In-D data varied. Our findings indicated that combining OOD detection with CRC in TRUECAM resulted in a robust empirical coverage that closely matched the target of 0.95 ($\alpha$=0.05) across a broad range of OOD-to-In-D ratios (Fig.~\ref{fig:SNGP_CP_complementary_summary}d). Notably, as the ratio increased, where more OOD data were misclassified as In-D, we observed a corresponding increase in the prediction set size to accommodate these risks (i.e., triggering more abstention) for maintaining a valid empirical coverage. By contrast, removing either CRC or OOD detection module significantly harmed the model's empirical coverage, with the decline becoming more severe as the proportion of OOD data increased. 

In addition, we investigated if implementing an inspection procedure (as described in the OOD detection above) before applying trained models on a new population's data (CPTAC in our scenario) can address performance degradation (observed in Fig.~\ref{fig:cp_comparison}a-d) due to distribution shift between datasets, thereby creating a more reliable model transfer environment. This mechanism, which we call distribution shift control (DSC), aims to ensure that deploying a trained model in different settings, particularly those without fine-tuning capabilities, is done with sufficient safeguards. We observed that excluding the CPTAC data identified by DSC at an OOD score threshold with FPR=0.2 enhanced NSCLC subtyping classification performance (Fig.~\ref{fig:SNGP_CP_complementary_summary}e,f) and CP efficiency by reducing set sizes across all settings (Fig.~\ref{fig:SNGP_CP_complementary_summary}i,j). A sensitivity analysis demonstrated that using an OOD score threshold corresponding to a higher FPR (and consequently higher TPR) consistently enhanced NSCLC subtyping accuracy (Fig.~\ref{fig:SNGP_CP_complementary_summary}g), AUROC (Fig.~\ref{fig:SNGP_CP_complementary_summary}h), average set size (Extended Data Fig.~\ref{fig:SNGP_CP_complementary_summary_extended}a), and DA error rate (Extended Data Fig.~\ref{fig:SNGP_CP_complementary_summary_extended}b). Again, SNGP-EAT surpassed all the other models by a large margin across the considered metrics (Supplementary Tables~\ref{tab:dsc_classification_tile}-\ref{tab:dsc_CP_patient}). 

\begin{figure}[!ht]
    \centering
    \includegraphics[scale=0.24]{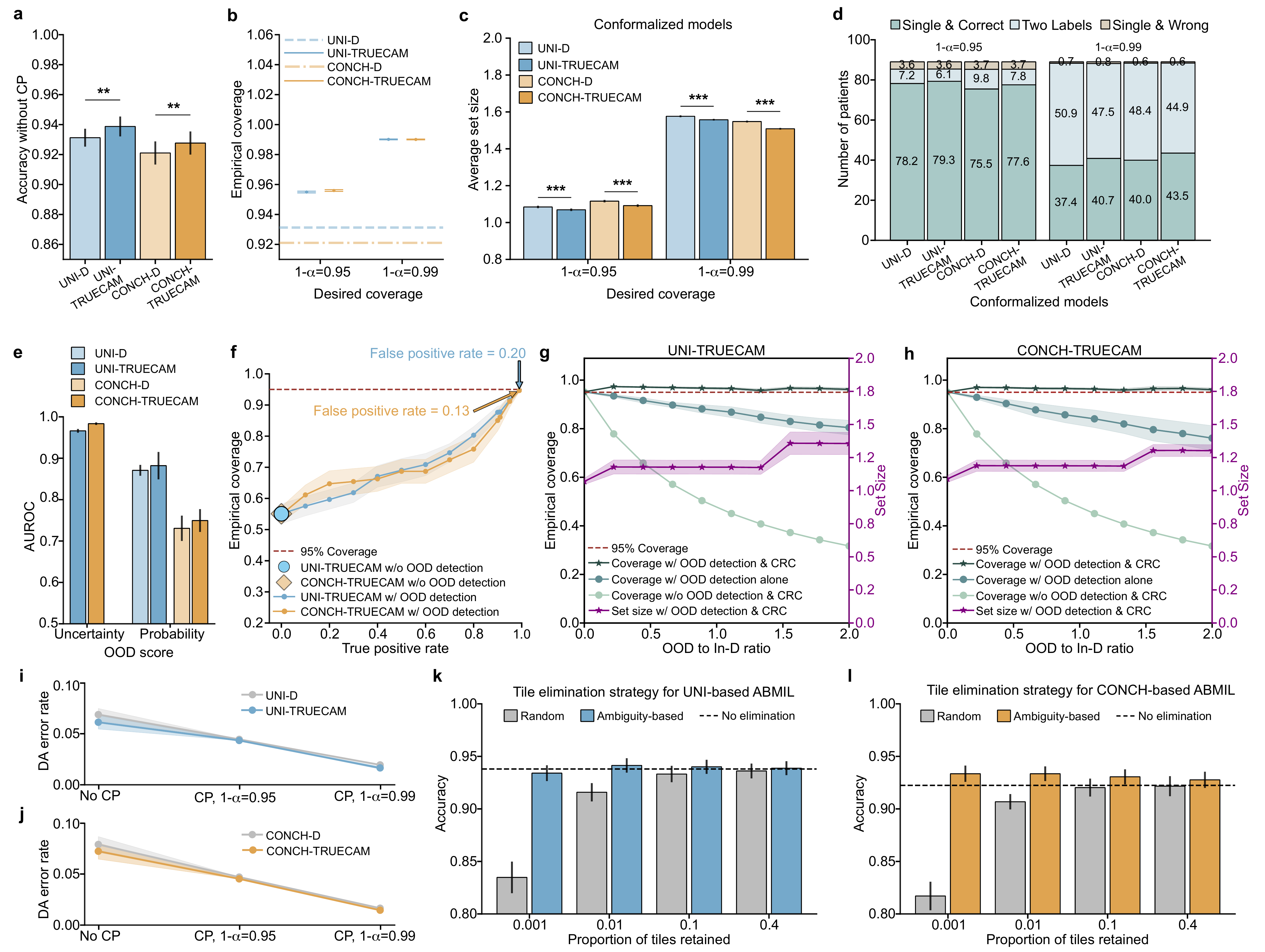}
    \caption{Assessment of TRUECAM using the foundation models UNI and CONCH. Results on TCGA (n=189) are reported. See Extended Data Fig.~\ref{fig:foundation_model_CPTAC_summary} for results on CPTAC, which produced similar observations. \textbf{a}, Classification accuracy of foundation model-based ABMIL with (denoted using suffix ``-TRUECAM'') and without TRUECAM (i.e., the original version of foundation model-based ABMIL, denoted using suffix ``-D''), evaluated without CP. \textbf{b}, Comparison of empirical coverage with and without TRUECAM. \textbf{c}, Comparison of average set size of CP with and without TRUECAM. \textbf{d}, Comparison of patient classification breakdown with and without TRUECAM. \textbf{e}, Performance of two distinct OOD scores in identifying slides outside a model’s scope. The uncertainty-based score is not applicable to the original versions of UNI and CONCH. \textbf{f}, Empirical coverage with $\alpha$ = 0.05 as a function of the true positive rate for OOD detection. \textbf{g}, \textbf{h}, Empirical coverage and set size of CP for UNI-TRUECAM and CONCH-TRUECAM under conformal risk control with a false positive rate of 0.20 for UNI and 0.13 for CONCH, shown as a function of OOD-to-In-D ratio. \textbf{i}, \textbf{j}, Comparison of patient-level DA error rate with and without TRUECAM in varying contexts of CP. \textbf{k}, \textbf{l}, Comparison of classification accuracy for UNI-TRUECAM and CONCH-TRUECAM using TRUECAM's ambiguity-based tile elimination strategy versus random tile elimination, shown as a function of the proportion of tiles retained per slide. One-sided Wilcoxon signed-rank test is utilized to calculate \textit{p} values. ** $p<0.01$, *** $p<0.001$. ABMIL, attention-based multiple instance learning; CP, conformal prediction; SNGP, spectral-normalized neural Gaussian process; EAT, elimination of ambiguous tiles; DA, definitive-answer.}
    \label{fig:foundation_model}
\end{figure}

\subsection*{TRUECAM's benefits extend to digital pathology foundation models}

Beyond specialized medical AI models like Inception-v3, which perform tile-level inference and aggregation, TRUECAM seamlessly integrates with general-purpose foundation models to make slide-level inference directly, without the need for tile-level result aggregation (Fig. \ref{fig:overview}b). Specifically, TRUECAM utilizes a foundation model’s pretrained encoder to extract tile representations from each slide, assigns ambiguity values to discard confusing tiles, and processes remaining tiles either through ABMIL for tile-level models or a slide encoder for slide-level models.
For evaluation, we systematically assessed TRUECAM’s performance with two recently released foundation models for digital pathology: UNI~\cite{chen2024towards}, a visual model, and CONCH~\cite{lu2024visual}, a visual-language model. For these two models, TRUECAM enforces distance-preserving transformations within the widely-adopted attention-based multiple instance learning (ABMIL) architecture~\cite{ilse2018attention} by incorporating spectral normalization in fully-connected layers. The final layer is updated with a Gaussian process to enable UQ. To assess tile ambiguity for EAT within the context of foundation models, a tile-level classifier aligned with the target task remains essential. Thus, we integrated TRUECAM with an AutoML system, AutoGluon~\cite{erickson2020autogluon}, to identify ambiguous tiles. Upon the completion of EAT, the trained ABMIL model was directly reused for inference without the need for retraining. For slide-level foundation models like Prov-GigaPath and TITAN, the adaptation of TRUECAM was slightly different. See the Methods for details.


We now show that TRUECAM established both data and model trustworthiness for NSCLC subtyping using digital pathology foundation models, addressing shortcomings of their original versions in deployment environments. We highlight the following key observations. First, when applied to foundation model-based ABMIL in the absence of CP, both UNI-TRUECAM and CONCH-TRUECAM, which removed 60.0\% of tiles per slide based on the ambiguity score defined in the Methods, enhanced the patient-level NSCLC subtyping accuracy 
(Fig.~\ref{fig:foundation_model}a). Specifically, UNI-TRUECAM and CONCH-TRUECAM reduced the error rate by 10.04\% and 8.48\%, respectively, than their deterministic counterparts (Fig.~\ref{fig:foundation_model}i,j). 
This highlights the combined effectiveness of SNGP and EAT in overcoming model limitations that contribute to suboptimal performance. See Supplementary Table~\ref{tab:foundation_models_SNGP_EAT} for TRUECAM's impact on the classification performance of TITAN and Prov-GigaPath. Second, with CP activated, the statistical guarantee of model coverage was established, as evidenced by the close alignment between desired and empirical coverage when applying TRUECAM (Fig.~\ref{fig:foundation_model}b). By comparison, the original foundation models paired with ABMIL for NSCLC subtyping achieved a coverage of only 0.931 (UNI-D) and 0.921 (CONCH-D) and generally lacked the flexibility to adjust to customized error levels. 
Third, when compared to the conformalized foundation model-based ABMIL (i.e., conformalized UNI-D and CONCH-D, without SNGP-EAT), TRUECAM displayed advantages by achieving smaller average set sizes (Fig.~\ref{fig:foundation_model}c) and producing a greater number of single and correct classifications (Fig.~\ref{fig:foundation_model}d). 
Fourth, TRUECAM enabled fairer or comparable classifications for both foundation model-based ABMIL setups compared to scenarios without TRUECAM in most cases (Extended Data Figs.~\ref{fig:extended_figure_UNI_fairness} and~\ref{fig:extended_figure_CONCH_fairness}).


In foundation model-based inference, TRUECAM identified and rejected OOD inputs with similar effectiveness as in the specialized model setting. Using the same OOD design and dataset as previously described, we observed that the uncertainty-based OOD score (AUROC of 0.967 for UNI-TRUECAM and 0.984 for CONCH-TRUECAM) consistently outperformed the probability-based score (AUROC of 0.882 for UNI-TRUECAM, and 0.750 for CONCH-TRUECAM) in distinguishing between In-D and OOD WSIs (Fig.~\ref{fig:foundation_model}e, Supplementary Table~\ref{tab:foundational_models_ood_summary}). Thus, the uncertainty-based OOD score was favored in the OOD evaluation. Notably, without OOD input identification, the empirical coverage of UNI-TRUECAM and CONCH-TRUECAM fell significantly below the desired coverage (Fig.~\ref{fig:foundation_model}f); however, TRUECAM's OOD detection capability assisted in achieving the desired coverage by adjusting the threshold for OOD score to approach a near perfect TPR. 
To manage undetected OOD inputs in environments with varying OOD-to-In-D ratios, TRUECAM integrates CRC with OOD detection to effectively address these inputs, achieving an empirical coverage that aligned closely with the desired level and maintaining consistency across a broad range of OOD-to-In-D ratios for both UNI-TRUECAM and CONCH-TURECAM (Fig.~\ref{fig:foundation_model}g,h). By contrast, neglecting either CRC or OOD detection invalidated empirical coverage. 

We further evaluated the effectiveness of ambiguity-based tile elimination in TRUECAM's EAT by comparing it to eliminating tiles uniformly at random and then conducted a sensitivity analysis to assess how the proportion of retained tiles influences classification accuracy. We observed that ambiguity-based tile elimination outperformed random tile elimination across a wide range of tile retaining rates for both foundation model-based ABMIL models (Fig.~\ref{fig:foundation_model}k,l). It is remarkable that when ambiguity-based EAT was applied aggressively, even as low as a 0.1\% tile retention rate, it did not compromise classification accuracy.
By contrast, random tile elimination was much more sensitive to the tile retaining rate, leading to significant degradation in accuracy when performed at a comparable retention rate. This suggested that the ambiguity-based EAT was highly effective as it kept the most informative tiles in slides for NSCLC subtyping.

\subsection*{TRUECAM delivers efficient interpretation and inference}

Compared to traditional approaches that rely solely on attention for interpretation, TRUECAM offers a diagnostic basis for model inference through two interpretability layers informed by its UQ module: 1) tile-level ambiguity scores, which assess each tile's potential to introduce confusion and help eliminate corresponding regions, and 2) a global attention map that shows the reliance of slide-level inference on each remaining tile after EAT.

To validate the diagnosis basis of TRUECAM across models, we randomly selected and visualize two WSIs from the TCGA testing cohort: one representing LUAD (Extended Data Fig.~\ref{fig:first_concatenated}) and one for LUSC (Extended Data Fig.~\ref{fig:second_concatenated}). A pathologist (F.N.), blinded to any model-derived information, received these WSIs to delineate tumor regions with significant diagnostic relevance for NSCLC subtyping. A random selection of tiles from various representative regions was also provided to the pathologist to assess their informativeness for distinguishing LUAD from LUSC. We observed that the regions with relatively low ambiguity scores closely aligned with the annotated areas by the pathologist, which facilitated their discrimination between LUAD and LUSC using established diagnostic criteria. Specifically, these low-ambiguity regions contained cancer epithelial areas, which were critical for distinguishing NSCLC subtypes. By contrast, regions with higher ambiguity scores lacked distinct morphological features of either NSCLC subtype. 
For example, tiles with blue borders from both LUAD and LUSC WSIs contained inflamed non-neoplastic lung parenchyma, areas of stroma, or necrosis, all of which were devoid of tumor-specific morphological hallmarks (Extended Data Fig.~\ref{fig:first_concatenated},~\ref{fig:second_concatenated}). Some regions appeared as poorly preserved or fields with artifacts, further undermining their utility in subtype discrimination.

After applying EAT, the morphological features in regions with high attention weights became the primary basis for the model’s classification. As confirmed by the pathologist, the high-attention regions highlighted in blue were generally consistent with the diagnostic criteria used to determine the NSCLC subtype. Tiles with red borders exemplified this alignment: LUAD-specific regions revealed glandular arrangements characteristic of adenocarcinoma, such as well-formed acinar patterns, while LUSC-specific regions displayed intercellular bridges, focal keratinization, and dense eosinophilic cytoplasm, all consistent with the morphological features used in pathologic classification. However, not all regions contributed equally to subtype discrimination. Regions lacking clear indicators for either subtype were recognized by TRUECAM with correspondingly low attention weights. For example, tiles with purple borders included stroma, areas of inflammation, areas of necrosis, and poorly differentiated regions, none of which provided clear discriminative clues for LUAD vs LUSC. 

We then quantified TRUECAM’s interpretability compared to traditional models by introducing \textit{attention efficiency}, which we defined as the fraction of high-attention tiles identified by the model that pathologists also consider highly relevant for subtyping decisions. This metric captures the precision of a model's attention in recognizing regions critical for accurate classification. TRUECAM consistently achieved an attention efficiency higher than traditional attention across all three model types (Extended Data Fig.~\ref{fig:attention_inference_efficiency}a,b). Thus, the two-level interpretability offered by TRUECAM can not only aid pathologists in identifying task-relevant regions for more reliable conclusions, but also serve as a valuable verification source to build trust in AI applications.

Beyond enhanced model interpretability, TRUECAM exhibited advantages in inference speed over traditional and other baseline uncertainty quantification methods. Multiple key findings were observed. First, Inception-v3 equipped with SNGP (denoted as Incep-SNGP) achieved nearly the same tile inference speed as its deterministic counterpart (Extended Data Fig.~\ref{fig:attention_inference_efficiency}c). Additionally, it was 5$\times$ faster than models using MC Dropout. This was primarily due to SNGP's single-forward pass mechanism. Second, the integration of SNGP introduced negligible computational burden on foundation models, partially attributed to their inherently slower encoding process. Third, while foundation models demonstrated significantly lower inference speeds compared to Inception-v3, they delivered superior classification performance (Supplementary Tables~\ref{tab:iacp}, \ref{tab:cptac-conch-varing-sn}). Fourth, EAT significantly boosted slide inference speed by excluding a significant proportion of ambiguous tiles, while simultaneously enhancing classification and CP performance (Extended Data Fig.~\ref{fig:attention_inference_efficiency}d). This inference efficiency gain positioned TRUECAM closer to practical deployment, particularly in time-sensitive and resource-limited medical settings. 

\section*{Discussions}

Trustworthiness of pathology AI is often compromised by model limitations in delivering reliable performance and quantifying uncertainty with statistical implication. Moreover, data challenges like noisy patches and the complexity of model deployment environment, including OOD data and distribution shifts, further undermine trustworthiness of pathology AI.
To address these issues, we developed TRUECAM, a framework designed to seamlessly integrate with models with various architectures, purposes, and complexities, to enable responsible and trustworthy AI-driven NSCLC subtyping in pathology. 

TRUECAM holds significant implications for the practical application of pathology AI models in real-world clinical settings. The most immediate benefit of TRUECAM lies in its significant reduction in the likelihood of AI models producing incorrect results. However, its impact extends beyond improving diagnostic accuracy; it redefines how pathology AI can be responsibly integrated into clinical workflows of cancer diagnostics. To manage the risk of erroneous patient-level diagnoses and ensure statistically guaranteed error bounds, TRUECAM adaptively defers uncertain or OOD cases to expert pathologists, fostering a collaborative decision-making process. Importantly, the framework tackles a fundamental challenge in pathology AI: inference data may experience distribution shifts, whether due to temporal changes at the same site, inter-site variability, or demographic and medical practice-based differences, all of which can undermine model trustworthiness. By addressing these issues, TRUECAM ensures that AI acts as a reliable assistant rather than an inflexible tool, which complements clinical workflows while mitigating the risks of automation bias. Critically, TRUECAM’s deferral mechanism is proactive, customizable, and strategic, abstaining less frequently than other models while achieving superior accuracy. This demonstrates the framework's ability to recognize and act on its own limitations, which confines AI usage within a trusted scope and reduces the cognitive load on pathologists, who can focus their expertise and effort on the most challenging cases ~\cite{cao2023efficient, liang2022advances, bernhardt2022active}. By integrating reliability, adaptability, and collaboration, TRUECAM establishes itself as a transformative step toward making AI a trusted partner for clinical decision-making.

TRUECAM not only confines diagnostics to the trusted scope of pathology AI but also lays a solid foundation for responsibly expanding this scope through a built-in awareness of its limitations. In a human-AI collaborative framework, expert pathologists can assign definitive labels to slides flagged as OOD or those where the model abstains from making a prediction. These newly labeled data points can then be incorporated into the uncertainty quantification process to enable iterative refinement of the model and its conformal predictor, progressively broadening the trusted scope of pathology AI in a controlled and reliable manner. Such a progressive approach outlines a responsible development paradigm, ensuring that pathology AI evolves in tandem with its real-world applications. As trustworthiness takes center stage in medical AI, the ability to systematically address foundational models' limitations in clinical settings becomes critical~\cite{hager2024evaluation,pfohl2024toolbox}. By enabling pathology AI to adapt and expand their capabilities through iterative feedback and refinement, TRUECAM provides key insights that can guide the development of next-generation foundation models in pathology, paving the way for their scalable and trustworthy deployment in clinical workflows.


While SNGP and EAT are designed to address data trustworthiness and CP (with conformal risk control) to assure model trustworthiness, these components do not function independently but instead provide complementary benefits. SNGP-EAT establishes an environment that upholds data exchangeability for CP to function and ensures that the input data is within the scope of the trained pathology AI models. Meanwhile, CP guarantees that classifications based on SNGP-EAT are statistically valid in containing the ground truth labels with a pre-specified coverage guarantee. The absence of either component significantly compromises NSCLC subtyping reliability. For example, without SNGP-EAT-based OOD detection, the empirical coverage of CP for both specialized and foundation model-based models dropped substantially to 0.477 
(compared to the desired coverage of 0.95), resulting in 52.3\% incorrect patient subtyping even with CP in place (Fig. \ref{fig:SNGP_CP_complementary_summary}d). While retaining SNGP and EAT alone (without CP) improved classification performance compared to baseline models (Fig. \ref{fig:EAT_summary}), it lacked the ability to identify challenging inputs and abstain from providing a diagnosis, nor can it establish bounded error rates to ensure guaranteed validity. The interplay between these components creates a robust framework that not only enhances individual performance but also addresses the broader challenges of real-world variability in pathology workflows.

Aligned with the data valuation and data-centric AI paradigm emphasized by the machine learning community~\cite{cao2023efficient, liang2022advances, bernhardt2022active}, EAT exemplifies the ``less data, better performance'' phenomenon and delivers transformative benefits for NSCLC subtyping in TRUECAM. While conceptually similar to the topK max-pooling operator used in MI-Zero~\cite{lu2023visual, lu2024visual}, which aggregates the most informative tile-level predictions, EAT takes this approach a step further. By filtering out ambiguous tiles and focusing on high-diagnostic-value regions, EAT creates a feedback loop that enhances model learning through reinforced supervisory signals from diagnostically valuable slide areas, leading to improved tile representations and amplifying both efficiency and accuracy. 

EAT’s advantages are manifold. First, it significantly reduces the computational burden for slide inference. For example, in the specialized model (Inception-v3), EAT decreased inference computational load by 66\% (Fig. \ref{fig:EAT_summary}a), while in the foundation model-based setting, this efficiency gain reached up to 1000$\times$ without compromising accuracy (Fig. \ref{fig:foundation_model}k,l). This efficiency lowers barriers to deploying advanced pathology AI models in resource-constrained settings, such as small clinics and rural hospitals, enabling broader access to AI-driven diagnostics. Second, EAT enhances interpretability by directing the model’s focus to diagnostically relevant areas, closely aligning its attention with pathologists' annotations (Extended Data Fig.~\ref{fig:first_concatenated},~\ref{fig:second_concatenated}). This not only reduces the cognitive load for pathologists when reviewing AI-generated results but also fosters greater trust in the model’s outputs, since its decision-making process becomes more transparent and clinically meaningful. From a data annotation perspective, EAT offers a scalable and efficient mechanism for filtering out low-value tiles, substantially reducing the reliance on labor-intensive and costly pixel-level annotations. This streamlined process saves time and resources while addressing the need for domain-specific expertise, making EAT a practical tool for advancing future model development. Third, EAT has significant implications for promoting equity by addressing demographic fairness gaps in both classification performance and CP set size, as seen with the original specialized and foundation model-based models. By eliminating highly ambiguous tiles from each slide, EAT mitigates biases that could arise when traditional pathology AI approaches process all tiles indiscriminately, which may inadvertently reinforce harmful disparities in NSCLC subtyping inference. These disparities could result in unequal treatment of patients from different racial or sexual backgrounds~\cite{chen2023algorithmic}. Importantly, EAT achieved these fairness improvements without requiring explicit constraints during model training, which highlights its potential to improve the equity and clinical reliability of pathology AI systems. Fourth, EAT also has far-reaching implications for the development of pathology AI foundation models, particularly in optimizing their pretraining process. The current standard self-supervised learning frameworks (e.g., DINOv2~\cite{oquab2024dinov2}) for image encoder pretraining treat all tiles in a slide equally. However, it is well-established that only cancerous regions are critical for clinically relevant downstream tasks. Similarly, in visual-language foundation models like CONCH or TITAN, a large portion of tiles in each slide are irrelevant to the diagnostic reports and contribute little to aligning text and image modalities in the model’s representation space. By applying EAT to exclude non-relevant tiles during pretraining, the process could become significantly more efficient by focusing on the meaningful cancerous areas. This would not only reduce the computational cost of pretraining but also enable the model to learn better representations, ultimately improving its performance across various downstream tasks.

For OOD detection, we observed that the uncertainty-based OOD score consistently outperformed the probability-based OOD score in foundation models  (Fig.~\ref{fig:foundation_model}e, Extended Data Fig.~\ref{fig:foundation_model_CPTAC_summary}e, Supplementary Table~\ref{tab:foundational_models_ood_summary}), but showed inferior performance compared to the probability-based OOD score in the specialized model, i.e., Inception-v3, (Fig.~\ref{fig:SNGP_CP_complementary_summary}a). This phenomenon can be in part attributed to the representational power of the underlying model. Since the uncertainty estimated by SNGP is essentially determined by the distance between testing and training data, models with stronger feature representation capabilities—such as foundation models, which achieve higher classification accuracy than Inception-v3—provide more discriminative and semantically rich dimensions to distinguish between In-D and OOD data in the feature space. This also aligns with findings reported by  existing research: distance-based methods can outperform probability-based methods in OOD detection tasks, particularly when the feature space is well-structured and the Gaussian assumption holds~\cite{liu2023simple}. On the other hand, the performance difference in OOD detection could be due to the distinct mechanisms of probability-based and uncertainty-based OOD score for OOD detection~\cite{liu2023simple}. In essence, for OOD detection, the probability-based score examines a data point's distance from the learned decision boundary in the feature space, while the uncertainty-based OOD score estimates the uncertainty of the data based on its distance from the training data. As foundation models learn better representations of training data than Inception-v3, the resultant representations enable fine-grained and informative estimates for the distance to the training data in the learned feature space. 


There are several limitations that we wish to highlight as opportunities for furthering this line of research. First, this study focused on the binary NSCLC subtyping problem of discriminating LUAD vs LUSC. However, the generalizability of TRUECAM's advantages in addressing broader multi-class cancer subtyping tasks in digital pathology needs to be assessed in settings with a larger number of tumor histologies, pathological patterns, and dataset characteristics. Second, we evaluated TRUECAM using one specialized model and four cutting-edge pathology AI foundation models that lack conversational interaction capabilities. Yet with the emergence of interactive pathology AI foundation models~\cite{moor2023foundation,huang2023visual,sun2024pathasst}, which lower the barrier to practical use by enabling conversational queries and reasoning regarding WSIs, it is important to investigate and assess how TRUECAM can be adapted to this new language-centered paradigm to further enhance model reliability. Third, while our findings suggested that distribution shift control enhanced the classification and CP performance on an external dataset (CPTAC) collected from a program distinct from the training dataset (TCGA) (Fig.~\ref{fig:SNGP_CP_complementary_summary}e,f,i,j), the root cause of the distribution shift remains unidentified. Further research is warranted to systematically investigate these distribution shifts and pinpoint their underlying cause, which could guide the model refinement.
Fourth, we did not evaluate the overall empirical performance of TRUECAM in a human-in-the-loop setting, particularly when abstention is triggered and pathologists are involved for independent inference.
In the future, well-designed trials will be critical to evaluate the impact of TRUECAM-assisted clinical decision-making for NSCLC subtyping. In practice, pathologists may choose to accept or override the definitive answers provided by error-bounded TRUECAM, influenced by factors such as the pathological features of specific slides and their own experience and confidence. The effect of these decisions on patient outcomes remains unclear. Therefore, it is essential to evaluate and explore effective human-AI collaboration strategies before deploying TRUECAM in clinical settings.

\section*{Method}
\subsection*{Dataset description}
\textbf{Dataset overview.} 
Supplementary Table~\ref{tab:data_summary} summarizes the basic characteristics of the datasets used in this study. We utilized two publicly available digital pathology datasets of NSCLC stained with hematoxylin and eosin (H\&E): 1) 941 WSIs from TCGA, comprising 467 LUAD and 474 LUSC cases, and 2) 1,306 WSIs from CPTAC, comprising 644 LUAD and 662 LUSC cases. In the TCGA dataset, each patient is represented by a single slide, whereas in the CPTAC dataset, each patient is associated with an average of 3.1 slides. Among the 416 patients in CPTAC, 203 are diagnosed with LUAD, while the remaining 213 are diagnosed with LUSC. 
To evaluate TRUECAM's ability in detecting OOD inputs, we constructed a TCGA-OOD dataset, consisting of 698 WSIs from 631 patients with non-lung, non-adenocarcinomas, and non-squamous tumors. TCGA-OOD includes slides from 92 patients with adrenocortical carcinoma (ACC), 70 patients with uveal melanoma (UVM), 57 patients with uterine carcinosarcoma (UCS), and 412 patients with bladder urothelial carcinoma (BLCA), representing all available WSIs from each considered cancer category in TCGA. Among these, except for UVM, one patient is associated with more than 1 tissue slides. We included BLCA, which is comprised of over 65\% of the dataset, because its urothelial origin and histological spectrum—ranging from low-grade papillary lesions to high-grade invasive forms—can closely resemble poorly differentiated NSCLC. In particular, high-grade BLCA exhibits certain morphological features that make it very challenging to distinguish from NSCLC, such as solid nests or sheets of atypical cells. As a result, TCGA-OOD serves as a meaningful dataset for the OOD detection task.

\textbf{Image preprocessing.} Regardless of distinct model input requirements, all WSIs were segmented and split into appropriate tiles using a standard preprocessing pipeline implemented in Slideflow (version 2.1.0)~\cite{dolezal2024slideflow}. Directly feeding large gigapixel WSIs into a neural network is computationally impractical, so image tiles of manageable sizes were extracted to align with the model's specifications. For example, foundation models like UNI, CONCH, and Prov-GigaPath accept image tiles of size $256 \times 256$ at 20X magnification; as we follow the Inception-v3 model developed in~\cite{dolezal2022uncertainty}, it requires image tiles of $299 \times 299$ at 10X magnification; unlike UNI, CONCH, and Prov-GigaPath, the latest foundation model TITAN accepts image tiles with a size of $512 \times 512$ at 20X magnification. The preprocessing pipeline also includes several standard steps to create deep learning-ready datasets. Otsu’s thresholding~\cite{otsu1975threshold} was applied to differentiate the foreground (tissue) from the background (empty slide), followed by grayspace filtering to remove background tiles. All extracted tiles were then stain-normalized and saved in TFRecord format for model training, inference, and performance evaluation. Importantly, no pathologist-annotated regions of interest were used; instead, we relied solely on slide-level diagnoses and assigned a uniform weak label to all tiles extracted from a given WSI when needed for model training. 


\textbf{Dataset configuration.}
In the setup for the specialized model based on Inception-v3, we randomly partitioned the TCGA dataset, allocating 65\% of the patients' data for training, 15\% for validation, and 20\% for calibration and testing. 
We repeated the random partitioning procedure 20 times, producing 20 independently trained models for evaluation.
Within the 20\% allocated for calibration and testing, data from 100 patients was used for calibration to establish CP (referred to as the TCGA calibration dataset), while data from 89 patients was set aside for testing (referred to as the TCGA testing dataset). To assess the variability of conformal prediction performance, the 20\% split was randomized 500 times, generating distinct calibration and testing datasets for each model. As a result, we report model performance based on 20 models $\times$ 500 evaluations per model. The CPTAC dataset was fully utilized for external validation to assess the effectiveness of TRUECAM in scenarios that need transferability, with data from 100 patients designated for calibration to establish conformal prediction (referred to as the external CPTAC calibration dataset) and the remaining 316 patients reserved for testing (referred to as the external CPTAC testing dataset). Similarly, the CPTAC dataset was randomly partitioned 500 times to evaluate the variability of conformal prediction performance.


In the scenario of using foundation models for NSCLC subtyping, which involves training a separate model using latent tile representations generated by foundation models, the TCGA and CPTAC datasets were utilized independently for the entire model training and evaluation process. Each dataset was partitioned into three subsets: 65\% for training, 15\% for validation, and 20\% for calibration and testing. Consistent with the strategy used for Inception-v3 evaluation, this partitioning and model training process was repeated 20 times, with each involving 500 random splits to establish and evaluate CP. 



\subsection*{Spectral-normalized Neural Gaussian Process (SNGP)}\label{subsec:SNGP}
We leveraged SNGP ~\citep{liu2023simple,liu2020simple} to provide a principled estimation of data uncertainty in deep learning models for NSCLC subtyping. SNGP introduces two important modifications to a regular neural network: 1) incorporating spectral normalization (SN) into the hidden layers and 2) replacing the conventional dense output layer with a Gaussian process (GP). Incorporating SN ensures that the transformation from the input $\bm{x}$ in the original input space to the latent representation in the penultimate layer $h \left(\bm{x} \right)$ preserves the relative distances. In other words, the distance $\left\| {h \left(\bm{x}_1 \right) - h \left(\bm{x}_2 \right)} \right\|_\mathrm{H}$ between representations of any two image data points $\left( {\bm{x}_1, \bm{x}_2} \right)$ in the latent space meaningfully reflects their their distance $\left\| {\bm{x}_1 - \bm{x}_2} \right\|_\mathrm{X}$ in the original data space, relative to other data pairs. In practice, adding SN addresses a problem in training neural networks known as feature collapse~\citep{van2020uncertainty}. This issue occurs when OOD data or data from different classes, which are geometrically distant from remaining data in the input space, are unexpectedly mapped to nearby points in the latent space, thus leading to unreasonable uncertainty estimations. Mathematically, the SN operation is equivalent to requiring the mapping function $h(\cdot)$ to satisfy the following bi-Lipschitz condition for any pair of inputs $\bm{x}_1$ and $\bm{x}_2$~\citep{o2006metric,liu2020simple}:
\begin{equation}
L_1 \times || \bm{x}_1 - \bm{x}_2 ||_\mathrm{X} \leq || h(\bm{x}_1) - h(\bm{x}_2) ||_\mathrm{H} \leq L_2 \times || \bm{x}_1 - \bm{x}_2 ||_\mathrm{X}, 
\label{eq:bi_Lipschitz_constraint}
\end{equation}
where $L_1$ and $L_2$ are the positive lower and upper Lipschitz bounds ($0 < L_1 < 1 < L_2 $) of feature extractor $h(\cdot)$, and $||\cdot ||_\mathrm{X}$ and $||\cdot ||_\mathrm{H}$ correspond to the distance metrics chosen for the original and latent space, respectively. The lower Lipschitz bound $L_1$ serves to preserve distances in the latent space and the upper Lipschitz bound $L_2$ helps enforce the smoothness and robustness of the neural network by limiting over-sensitive transformations to perturbations in the input space of $\bm{x}$.


In SNGP, the bi-Lipschitz constraint is enforced by applying spectral normalization to neural network weights $\left\{ {{\bm{W}_l}} \right\}_{l = 1}^{L - 1}$~\citep{behrmann2019invertible}, where $l$ represents the $l$-th hidden layer of the neural network. During each training step, the SN method first estimates the spectral norm $\widehat \lambda  \approx {\left\| {{\bm{W}_l}} \right\|_2}$ using the power iteration method~\citep{gouk2021regularisation}, and then normalizes the neural network weights as:
\begin{equation}
{\widehat {\bm{W}}_l} = \left\{ \begin{array}{l}
c\bm{W}_l/\widehat \lambda,\quad \text{if}\;c < \widehat \lambda \\
{\bm{W}_l},\quad \text{otherwise}
\end{array} \right.,
\end{equation}
where $c$ is a hyperparameter used to adjust the exact spectral norm upper bound on ${\left\| \bm{W}_l \right\|_2}$ (i.e., ${\left\| \bm{W}_l \right\|_2} \le c $). 

SNGP then estimates data uncertainty by leveraging the preserved distances in the latent space, achieved by replacing the dense output layer with a GP. Consider a dataset consisting of $N$ training points $\mathcal{D} = \left\{ {\bm{x}_i,y_i} \right\}_{i = 1}^N$, and let $h_i(\bm{x}_i)$ denote the representation of $\bm{x}_i$ in the second-to-last layer of neural network. The output $\bm{f}_{N\times1} = \left[ {f(h_1), f(h_2), \cdots, f(h_N)} \right]^T$ of GP follows a multivariate Gaussian distribution:
\begin{equation}\label{eq:GP_prior}
    \bm{f}_{N\times1} \sim \mathcal{\bm{N}} \left(\bm{0}_{N\times 1}, \bm{K}_{N\times N} \right), \text{where} \;\; \bm{K}_{i,j} = \exp \left( {\frac{{ - \left\| {h_i - h_j} \right\|_2^2}}{2}} \right).
\end{equation}

Conditional on the learned latent representation $h_i$, SNGP approximates GP using the Laplace approximation with the random Fourier feature (RFF) expansion~\citep{williams2006gaussian}. Specifically, SNGP approximates the GP prior in Eq. (\ref{eq:GP_prior}) with a low-rank approximation to the kernel matrix $\bm{K} = \Phi \Phi^T $ using random features~\citep{rahimi2007random}: 
\begin{equation}
    \bm{f}_{N\times1} \sim \mathcal{\bm{N}} \left(\bm{0}_{N\times 1}, \Phi\Phi^T_{N\times N} \right), \text{where} \;\; \Phi_{i} = \sqrt {\frac{2}{D_L}} \cos \left( { - \bm{W}_Lh_i + \bm{b}_L} \right),
\end{equation} 
where $\Phi_i$ is the last layer of the neural network with a dimension $D_L$, $\bm{W}_L$ is a fixed weight matrix randomly generated from $\mathcal{N}(0, 1)$ and $\bm{b}_L$ is a fixed bias vector randomly generated from a uniform distribution $\mathcal{U}(0, 2\pi)$.  
Since $\bm{W}_L$ and $\bm{b}_L$ are fixed, the RFF approximation to the GP prior for the $k$-th logit of the classification problem (in Eq. (\ref{eq:GP_prior})) can be reformulated as:
\begin{equation}
    g_k\left( {{h_i}} \right) = \sqrt {\frac{2}{{{D_L}}}} \cos \left( { - \bm{W}_Lh_i + \bm{b}_L} \right)^T\times \beta_k, \quad k = 1, \cdots, K,
\end{equation}
where $\beta_k \sim \mathcal{N}\left(0, \bm{I} \right)$, $K$ denotes the number of classes in the classification problem, and $\bm{\beta} = \left\{ {\beta_k}\right\}_{k=1}^{K} $ are the only learnable parameters in the output layer.

Overall, the trainable parameters consist of $\bm{\beta}$ along with the weights and biases $\bm{\alpha} = \left\{ {\bm{W}_l, \bm{b}_l}\right\}_{l=1}^{{L-1}} $ in the first $L-1$ layers of the neural network. Notably, the RFF approximation to GP reduces the high-dimensional GP to a standard Bayesian linear model while also producing a closed-form posterior that is end-to-end trainable alongside the rest of the neural network using stochastic gradient descent. 
After inferring the MAP estimate of $\hat{\beta_k}$, for an unobserved point $\bm{x}^*$, the first $L-1$ layers of SNGP serve as a feature extractor to derive its latent representation $\Phi^* = \sqrt {\frac{2}{{{D_L}}}} \cos \left( { - \bm{W}_Lh(\bm{x}^*) + \bm{b}_L} \right)$. Next, the mean and variance associated with the $k$-th logit in the classification problem are computed as:
\begin{equation}
\begin{array}{l}
    \mu_k \left(\bm{x}^* \right) = {\left[ {\Phi^*} \right]^T\hat{\beta_k}}, \quad
    \sigma_k \left(\bm{x}^* \right) = \left[ {\Phi^*}\right]^T \Sigma_k \Phi^*.
\end{array}
\end{equation}

Finally, the predictive distribution for $\bm{x}^*$ is calculated as:
\begin{equation}
    p\left( {\left. {y^*} \right|\bm{x}^*} \right) = \int\limits_{s \sim \mathcal{N} (\mu_k \left(\bm{x}^* \right), \sigma_k \left(\bm{x}^* \right))} {\gamma (s)},
\end{equation}
where $\gamma$ denotes the softmax activation function.


\subsection*{Conformal prediction}
Conformal prediction, or CP, is a distribution-free approach for constructing prediction sets for arbitrary prediction or classification algorithm. It provides statistical guarantees of a bounded error rate for model outputs, rigorously aligned with the error level specified by developers or service providers under data exchangeability assumption~\citep{angelopoulos2023prediction}. Consider a classification problem comprising of $K$ classes. Let $\mathcal{D} = \left\{ {\bm{x}_i,y_i} \right\}_{i = 1}^T$ be independently and identically distributed (i.i.d.) data points sampled from distribution $\mathcal{X} \times \mathcal{Y}$, where $\bm{x}_i \sim \mathcal{X}$, $y_i \sim \mathcal{Y}$, $y_i \in  \{1, \cdots, K \} $, and $\alpha \in \left[ {0, 1} \right]$ denotes a specified error level (also referred to as significance level). Then the objective of CP is to construct a prediction set for each data point, denoted as
\begin{equation}
    \Gamma: \mathcal{X} \to \left\{ {\text{subset of} \;\; \left\{ {1,2, \cdots ,K} \right\}} \right\},
\end{equation}
such that for a new i.i.d. data point $(\bm{x}^*, y^*) \sim P\left(\mathcal{X}, \mathcal{Y} \right)$, the probability that the prediction set fails to contain the truth label is $\alpha$ bounded. This is formally expressed as~\cite{lei2018distribution}:
\begin{equation}\label{eq:cp_constraint}
   1-\alpha  \le \mathbb{P}\left( {y^* \in \Gamma (\bm{x}^*) } | \bm{x}^*) \right) \le  1-\alpha + \frac{1}{T+1},
\end{equation}
where the probability is taken over the data $(\bm{x}_i, y_i)_{i=1}^{T} \cup (\bm{x}^*, y^*)$.
By design, CP guarantees that the prediction set $\Gamma (\bm{x}^*)$ contains the true label $y^*$ with a probability of at least $1 - \alpha$. 
The bounded error rate is achieved by generating a set of likely class labels for the inputs with insufficient certainty in a systematic way, in contrast to the single-value (in regression) or single-label (in classification) outputs produced by regular machine learning models. 


Using two separate and disjoint datasets for model training and calibration, the development of an inductive conformal predictor consists of three key steps:
\begin{enumerate}
    \item \textbf{Define a nonconformity score}. Conformity reflects how similar or conformal a new data point is to the training data~\citep{angelopoulos2021gentle} and is typically quantified using a nonconformity score. In classification tasks, the nonconformity score is oftentimes defined as $s(\bm{x}_i, y_i) = 1-\widehat{f}\left(\bm{x}_i \right)_{y_i}$, where $\widehat{f}\left(\bm{x}_i \right)_{y_i}$ denotes the probability assigned by the trained model $\widehat{f}$ to the true class $y_i$ for the input $\bm{x}_i$. Thus, the lower the model's confidence in correctly classifying $\bm{x}_i$ as $y_i$, the higher the corresponding nonconformity score. We used this nonconformity score throughout the study.
    
    \item \textbf{Determine the nonconformity threshold}. We then computed the nonconformity score $s\left(\bm{x}_i, y_i \right)$ for each data point $\bm{x}_i$ ($i=T+1, \cdots, T+R$) in the calibration set and obtain a set of nonconformity scores: $\{s(\bm{x}_{T+1}, y_{T+1}), s(\bm{x}_{T+2},  y_{T+2}), \allowbreak \cdots, s(\bm{x}_{T+R}, y_{T+R})\}$. We ranked these scores in ascending order and determine the threshold $\widehat q$ as the $\frac{\left\lceil {(R+1)(1-\alpha)} \right\rceil}{R}$ quantile of these ordered scores. This threshold ensures that the resulting prediction sets have a coverage rate of at least $1-\alpha$ on new i.i.d. datasets.

    \item \textbf{Generate prediction sets}. For a new data point $\bm{x}^*$, the prediction set is formed by including all class labels $k\in \{1,2,...,K\}$ whose corresponding scores $s\left( \bm{x}^*, k \right)$ are smaller than or equal to $\widehat{q}$. Mathematically, we have:
    \begin{equation}
        \Gamma (\bm{x}^*) = \left\{ {{k:s\left( \bm{x}^*, k \right) \le \widehat q} } \right\}, \;  k = 1, \ldots , K.
    \end{equation}
\end{enumerate}

Unlike regular uncertainty quantification methods, CP converts the traditional heuristic concept of uncertainty, often represented by softmax outputs, into a statistically meaningful measure in the form of a prediction set. By following the steps described above, CP ensures to generate statistically valid prediction sets that contain the true label with a pre-specified coverage of $1 - \alpha$. This approach offers a standardized framework for constructing reliable confidence intervals that enables machine learning systems to deliver robust outputs while upholding responsible and trustworthy scientific inferences.


\subsection*{Specialized models based on Inception-v3}
We trained three deep learning models based on the Inception-v3 architecture ~\cite{szegedy2015going, kers2022deep}. The first is the original deterministic version, which does not include UQ. The other two models do incorporate UQ techniques: 1) MC Dropout~\cite{gal2016dropout} and 2) SNGP. Inception-v3 is a 48-layer deep convolutional neural network designed to capture spatial features at varying scales by employing parallel convolution operations with different filter sizes within the same layer. To address the vanishing gradient problem and provide additional supervision signals, Inception-v3 integrates an auxiliary classifier to propagate label information to earlier network layers. In addition, Inception-v3 utilizes bottleneck layers to reduce dimensionality and computational costs. These design features collectively enhance performance and resource efficiency, rendering Inception-v3 widely adopted for a diverse range of general computer vision and medical imaging applications. Supplementary Table \ref{tab:inception_parameter_summary} provides parameter and training details.
The integration of MC dropout with Inception-v3 follows the design adopted by Dolezal et al.~\cite{dolezal2022uncertainty}. At the inference stage, uncertainty is estimated by generating an ensemble of predictions, where each image tile is passed through the dropout-enabled Inception-v3 five times. 

We trained models from scratch using a 24GB NVIDIA RTX 4090 GPU with a batch size of 64, where image tiles are resized to $299 \times 299$ pixels. We adopted a learning rate scheduler with an initial learning rate of 0.0003 and a decay rate of 0.98 every 512 steps. Each model is trained for four epochs using the Adam optimizer with the 1st and 2nd moment exponential decay rate setting to 0.9 and 0.999, respectively. During model training, we monitored model performance on the validation dataset and save the best-performing model for evaluation on the testing data. Standard data augmentation techniques, including random horizontal and vertical flips, 90-degree rotations, JPEG compression with a 50\% probability (quality range: 50-100), and Gaussian blur with a 10\% probability (sigma: 0.5-2.0), are used to augment the training process of each model.

\subsection*{Foundation models}
TRUECAM seamlessly integrates with digital pathology foundation models for trustworthy NSCLC subtyping. In this study, we explore integrating TRUECAM with two types of foundation models: tile-level (or patch-level) models, including UNI~\cite{chen2024towards} and CONCH~\cite{lu2024visual}, and slide-level models, including Prov-GigaPath~\cite{xu2024whole} and TITAN~\cite{ding2024multimodal}, both using the vision transformer-based architecture as imagine encoders. All of them are general-purpose models systematically pretrained on diverse sources of histopathology images using self-supervised strategies. Among them, CONCH, Prov-GigaPath, and TITAN are visual-language models designed to align image and text modalities within their representation space, whereas UNI is exclusively image-pretrained. These foundation models exhibit exceptional adaptability and knowledge transferability through image encoding, making them highly effective for a wide range of downstream tasks in histopathology, including image classification, region-of-interest retrieval, and cell type segmentation.


To adapt tile-level foundation models for NSCLC subtyping, we utilized the image encoders of UNI and CONCH to extract tile embedding (i.e., latent representations), concatenate the resulting representations for each slide, and apply the widely used attention-based multiple instance learning (ABMIL) for weakly supervised classification using slide-level diagnostic labels. The ABMIL model architecture and training configuration follow those established in UNI~\cite{chen2024towards} and CONCH~\cite{lu2024visual}. Specifically, the ABMIL architecture includes a fully-connected layer with a rectified linear unit (ReLU) activation function to transform tile-level features into 512-dimensional embedding, followed by an intermediate layer with size 384. Finally, another fully-connected layer maps the attention-pooled slide-level representations to logits and class probabilities through softmax normalization. Dropout is applied at multiple stages: a rate of 0.1 for the input features and 0.25 after each intermediate layer. To incorporate SNGP, we added spectral normalization to all fully-connected layers and replace the final layer with a RFF-approximated Gaussian process. The ABMIL model is trained via a maximum of 20 epochs using the AdamW optimizer~\cite{loshchilovdecoupled} with a cosine learning rate scheduler, an initial learning rate of $1 \times 10^{-4}$, and cross-entropy loss.

For slide-level foundation models, we fine-tuned the slide encoders of Prov-GigaPath and performed linear probing on the slide representations of TITAN for downstream classification tasks as suggested by~\cite{ding2024multimodal} and~\cite{xu2024whole}. For both models,
we appended a Gaussian process layer to the slide-level representations without applying spectral normalization, as the representations extracted from Prov-GigaPath and TITAN inherently preserve distances effectively, due to their dedicated self-supervised training strategy.

Before the trained ABMIL models with SNGP and fine-tuned slide encoders with GP are utilized for NSCLC subtyping, EAT is applied to identify and remove ambiguous tiles from each slide. As the tile-level features extracted from foundation models are incompatible with the Inception-v3 architecture, we employed AutoGluon to train a machine learning model as a proxy for estimating ambiguity scores defined later. AutoGluon treats the extracted features as tabular data and searches the design space comprising multiple state-of-the-art models (e.g., LightGBM, NeuralNet, XGBoost, ExtraTrees) for NSCLC subtyping. It then automatically ensembles the weights of these base models to maximize the model performance on the validation data. Only the tiles with ambiguity score less than a given threshold are kept in the attention-pooled slide-level representations for NSCLC subtyping. Such a strategy eliminates 60.0\% tiles for TCGA and CPTAC, a level comparable to that achieved in the specialized model setting. 

\subsection*{Ambiguity score}
Regardless of the underlying tile-level classification model, given a trained classification model $M$, we define ambiguity score of a tile $\bm{x}$ as $s\left( {\bm{x},M} \right) = 1 -  \left| {p\left( {y(\bm{x}) = 0\left| {M} \right.} \right) - p\left( {y(\bm{x}) = 1\left| {M} \right.} \right)} \right|$, where $\left| {p\left( {y(\bm{x}) = 0\left| {M} \right.} \right) - p\left( {y(\bm{x}) = 1\left| {M} \right.} \right)} \right|$ measures how discernible tile $\bm{x}$ is by the model $M$. Clearly, the higher the ambiguity score, the lower $\left| {p\left( {y(\bm{x}) = 0\left| {M} \right.} \right) - p\left( {y(\bm{x}) = 1\left| {M} \right.} \right)} \right|$ and the less informative tile $\bm{x}$ to the model $M$ in the NSCLC subtyping task.
For Inception-v3, the model itself was used to calculate the ambiguity score. In contrast, for UNI, CONCH, and TITAN, their respective tile encoders were used to extract tile representations, and AutoGluon was trained from scratch to derive tile-level ambiguity scores. Prov-GigaPath utilized the AutoGluon model derived from CONCH to evaluate its transferability.


\subsection*{Uncertainty quantification}
MC Dropout is employed as the primary baseline approach for uncertainty quantification~\citep{gal2016dropout}.
By assuming that the prior of neural network weights follows a standard normal distribution, MC Dropout is mathematically equivalent to sampling from a variational posterior consisting of two independent Gaussians with fixed covariance. To implement MC Dropout in Inception-v3, we appended two fully-connected layers, each containing 1,024 neurons, to the end of the neural network and applied dropout with a rate of 0.1 after each fully-connected layer. During inference, each image tile was forward-passed through the dropout-enabled network five times. The standard deviation of the model's predictions across these five passes was calculated as the measure of tile-level uncertainty. The tile-level predictions and their corresponding uncertainty were then aggregated to produce slide- and patient-level predictions, along with the associated uncertainty estimates.

To derive data uncertainty in the context of SNGP, once the model is properly trained, the random Fourier features for $\bm{x}^*$ can be calculated as $\Phi^* = \sqrt {\frac{2}{{{D_L}}}} \cos \left( { - \bm{W}_Lh_i(\bm{x}^*) + \bm{b}_L} \right)$. With this $D_L$-dimensional latent representation of $\bm{x}^*$, the tile-level uncertainty estimates can be derived as $\sqrt {\tau \left[\bm{\Phi}^*\right]^T (\bm{\Phi}_{N \times D_L}^T \bm{\Phi}_{N \times D_L} + \tau \bm{I})^{-1}\bm{\Phi}^*}$, where $\tau$ denotes the ridge factor. SNGP requires only a single forward pass of an image tile through the neural network to extract the corresponding RFF features for uncertainty estimation. By contrast, MC Dropout requires multiple forward passes of the same image tile to quantify prediction uncertainty. This makes SNGP more computationally efficient during inference, which is particularly important for large-scale, safety-critical applications with stringent real-time inference requirements like cancer diagnostics.

It is important to note that SNGP and CP differ in their approaches for uncertainty quantification. Beyond the regular classification probability of $\bm{x}^*$ produced by a deep learning model, SNGP generates an estimated standard deviation $\sigma(\bm{x}^*)$ as the uncertainty measure for $\bm{x}^*$. By contrast, CP does not produce any explicit uncertainty measure. Instead, it operates within the class probability space established by the deep learning model and constructs a prediction set $\Gamma (\bm{x}^*)$ for $\bm{x}^*$ by including labels with the corresponding prediction probabilities beyond a threshold. Thus, SNGP and CP fundamentally differ in how they quantify uncertainty. 
 
For OOD detection and DSC, two scores were considered: 1) a probability-based OOD score, and 2) an uncertainty-based OOD score. The probability-based OOD score aggregates tile-level class probabilities and measures the degree of OOD as $1 - \frac{1}{N} \sum_{i=1}^{N} \max_{k} p(\hat{y}_k|\bm{x}_i)$, where $p(\hat{y}_k|\bm{x}_i)$ denotes the probability of tile $i$ being classified as $\hat{y}_k$. Unlike the probability-based OOD score, the uncertainty-based OOD score quantifies tile-level prediction uncertainty using the standard deviation estimated by MC Dropout or SNGP. It then employs the first $\delta$ tiles with the lowest  uncertainty for each patient to represent the patient-level prediction uncertainty. Clearly, the uncertainty-based OOD score offers additional insight beyond what  deterministic models can provide, effectively highlighting the inherent decision limitations of deep learning models, as demonstrated by its superior performance in identifying OOD images in the foundation model-based approach. 

\subsection*{Conformal risk control (CRC)}
CRC extends regular CP to accommodate a broader range of loss functions that assess different aspects of prediction quality beyond miscoverage, such as accuracy, calibration, and other task-specific performance metrics~\cite{angelopoulos2022conformal}. 
More concretely, we define the prediction set for $\bm{x}$ by model $\widehat{f}$ as ${\Gamma _\rho }\left( {\bm{x}} \right) = \{ k:\widehat{f}{\left( \bm{x} \right)_{k}} \ge 1 - \rho \}$, where $k\in \{1, \cdots , K\}$ denotes all possible class labels and $\rho$ is a threshold that controls the conservativeness of the prediction set. Then, for any bounded loss function $l(\cdot)$ that is non-increasing as the prediction set ${\Gamma _\rho }\left( {\bm{x}} \right)$ gets larger, CRC provides a statistical guarantee on the loss in the following form:
\begin{equation}\label{eq:conformal_risk_control}
    \mathbb{E} \left[ {l\left( {{\Gamma _\rho }\left( {{\bm{x}}} \right),{y}} \right)} \right] \le \alpha.
\end{equation}

Compared to regular CP, CRC extends CP to control the expected value of any monotone loss function instead of only coverage. The goal of CRC is to determine the lowest possible $\widehat {\rho}$ based on the calibration dataset, such that the risk control defined in Eq.~(\ref{eq:conformal_risk_control}) holds for unseen data.
In the deployment of pathology AI where In-D and OOD data coexist, the observed coverage over the calibration data guides the identification of $\widehat {\rho}$  via binary search. The threshold $\widehat {\rho}$ can be iteratively adjusted until the target coverage $1 - \alpha$ is satisfied in the presence of OOD data. After the optimal $\widehat {\rho}$ is identified, CRC then applies it to the unseen data such that their coverage is $1 - \alpha$ guaranteed. 

\subsection*{Model fairness evaluation}
For model fairness evaluation, we examined the performance gap between the best- and worst-performing subgroups based on sex and race~\cite{ktena2024generative}. Two performance dimensions were assessed: 1) accuracy of model predictions, and 2) size of CP prediction sets. Group-level average values of these metrics were calculated to represent the performance of each corresponding subgroup. Since the gap value is always non-negative, a smaller gap indicates a fairer model.
Sex and race information is available in both the TCGA and CPTAC datasets. TCGA includes 569 male and 372 female patients, while CPTAC comprises 288 male and 127 female patients, with one patient’s sex not reported. For race, categories with fewer than 20 patients are grouped as ‘Others’. Specifically, in TCGA, patients are grouped into Others (n=98), White (n=678), and Not reported (n=165). In CPTAC, patients are grouped into Asian (n=123), Others (n=49), and While (n=244). Demographic information is summarized in Supplementary Table~\ref{tab:data_summary}.

\section*{Data Availability}
The TCGA diagnostic whole-slide data and corresponding labels are available from the NIH genomic data commons (\url{https://portal.gdc.cancer.gov}). The CPTAC whole-slide data and the corresponding labels are available from the NIH cancer imaging archive (\url{https://cancerimagingarchive.net/datascope/cptac}). We summarized the links to all the data used in this paper in Supplementary Table~\ref{tab:summarized_links_to_data}.

\section*{Code availability}
All code was implemented in Python using PyTorch as the primary deep-learning library. The complete pipeline for processing WSIs as well as training and evaluating models is available at \url{https://github.com/iamownt/TRUECAM} and can be used to reproduce the experiments of this paper.

\section*{Acknowledgments}
X.Z. and T.W. were partially supported by a grant from the Research Grants Council of the Hong Kong Special Administrative Region, China (Project No. PolyU 25206422) and the National Natural Science Foundation of China (Grant No. 62406269). C.Y. was supported by the U.S. National Institutes of Health (NIH) NLM Pathway to Independence Award 1K99LM014428-01A1.

\section*{Author contributions statement}
X.Z., T.W., and C.Y. conceived the study and designed the experiments, X.Z. carried out data collection, T.W. wrote the codes, performed the experiments, and analyzed the experimental results. X.Z. and C.Y. wrote the original manuscript and summarized major experimental findings, Y.C. and B.M. reviewed and revised the manuscript, and F.N. annotated the two WSI slides and added descriptions on the role of different tiles for NSCLC subtyping. All authors contributed to manuscript preparation and approved the final version.

\bibliography{ref}

\begin{thebibliography}{10}
\urlstyle{rm}
\expandafter\ifx\csname url\endcsname\relax
  \def\url#1{\texttt{#1}}\fi
\expandafter\ifx\csname urlprefix\endcsname\relax\def\urlprefix{URL }\fi
\expandafter\ifx\csname doiprefix\endcsname\relax\def\doiprefix{DOI: }\fi
\providecommand{\bibinfo}[2]{#2}
\providecommand{\eprint}[2][]{\url{#2}}

\bibitem{bera2019artificial}
\bibinfo{author}{Bera, K.}, \bibinfo{author}{Schalper, K.~A.}, \bibinfo{author}{Rimm, D.~L.}, \bibinfo{author}{Velcheti, V.} \& \bibinfo{author}{Madabhushi, A.}
\newblock \bibinfo{journal}{\bibinfo{title}{Artificial intelligence in digital pathology—new tools for diagnosis and precision oncology}}.
\newblock {\emph{\JournalTitle{Nature Reviews Clinical Oncology}}} \textbf{\bibinfo{volume}{16}}, \bibinfo{pages}{703--715} (\bibinfo{year}{2019}).

\bibitem{rajpurkar2022ai}
\bibinfo{author}{Rajpurkar, P.}, \bibinfo{author}{Chen, E.}, \bibinfo{author}{Banerjee, O.} \& \bibinfo{author}{Topol, E.~J.}
\newblock \bibinfo{journal}{\bibinfo{title}{{AI} in health and medicine}}.
\newblock {\emph{\JournalTitle{Nature Medicine}}} \textbf{\bibinfo{volume}{28}}, \bibinfo{pages}{31--38} (\bibinfo{year}{2022}).

\bibitem{carrillo2024generation}
\bibinfo{author}{Carrillo-Perez, F.} \emph{et~al.}
\newblock \bibinfo{journal}{\bibinfo{title}{Generation of synthetic whole-slide image tiles of tumours from rna-sequencing data via cascaded diffusion models}}.
\newblock {\emph{\JournalTitle{Nature Biomedical Engineering}}} \bibinfo{pages}{1--13} (\bibinfo{year}{2024}).

\bibitem{thirunavukarasu2023large}
\bibinfo{author}{Thirunavukarasu, A.~J.} \emph{et~al.}
\newblock \bibinfo{journal}{\bibinfo{title}{Large language models in medicine}}.
\newblock {\emph{\JournalTitle{Nature Medicine}}} \textbf{\bibinfo{volume}{29}}, \bibinfo{pages}{1930--1940} (\bibinfo{year}{2023}).

\bibitem{dvijotham2023enhancing}
\bibinfo{author}{Dvijotham, K.} \emph{et~al.}
\newblock \bibinfo{journal}{\bibinfo{title}{Enhancing the reliability and accuracy of ai-enabled diagnosis via complementarity-driven deferral to clinicians}}.
\newblock {\emph{\JournalTitle{Nature Medicine}}} \textbf{\bibinfo{volume}{29}}, \bibinfo{pages}{1814--1820} (\bibinfo{year}{2023}).

\bibitem{begoli2019need}
\bibinfo{author}{Begoli, E.}, \bibinfo{author}{Bhattacharya, T.} \& \bibinfo{author}{Kusnezov, D.}
\newblock \bibinfo{journal}{\bibinfo{title}{The need for uncertainty quantification in machine-assisted medical decision making}}.
\newblock {\emph{\JournalTitle{Nature Machine Intelligence}}} \textbf{\bibinfo{volume}{1}}, \bibinfo{pages}{20--23} (\bibinfo{year}{2019}).

\bibitem{chua2023tackling}
\bibinfo{author}{Chua, M.} \emph{et~al.}
\newblock \bibinfo{journal}{\bibinfo{title}{Tackling prediction uncertainty in machine learning for healthcare}}.
\newblock {\emph{\JournalTitle{Nature Biomedical Engineering}}} \textbf{\bibinfo{volume}{7}}, \bibinfo{pages}{711--718} (\bibinfo{year}{2023}).

\bibitem{banerji2023clinical}
\bibinfo{author}{Banerji, C.~R.}, \bibinfo{author}{Chakraborti, T.}, \bibinfo{author}{Harbron, C.} \& \bibinfo{author}{MacArthur, B.~D.}
\newblock \bibinfo{journal}{\bibinfo{title}{Clinical {AI} tools must convey predictive uncertainty for each individual patient}}.
\newblock {\emph{\JournalTitle{Nature Medicine}}} \bibinfo{pages}{1--3} (\bibinfo{year}{2023}).

\bibitem{gal2016dropout}
\bibinfo{author}{Gal, Y.} \& \bibinfo{author}{Ghahramani, Z.}
\newblock \bibinfo{title}{Dropout as a {B}ayesian approximation: Representing model uncertainty in deep learning}.
\newblock In \emph{\bibinfo{booktitle}{International Conference on Machine Learning}}, \bibinfo{pages}{1050--1059} (\bibinfo{organization}{PMLR}, \bibinfo{year}{2016}).

\bibitem{abdar2022need}
\bibinfo{author}{Abdar, M.}, \bibinfo{author}{Khosravi, A.}, \bibinfo{author}{Islam, S. M.~S.}, \bibinfo{author}{Acharya, U.~R.} \& \bibinfo{author}{Vasilakos, A.~V.}
\newblock \bibinfo{journal}{\bibinfo{title}{The need for quantification of uncertainty in artificial intelligence for clinical data analysis: increasing the level of trust in the decision-making process}}.
\newblock {\emph{\JournalTitle{IEEE Systems, Man, and Cybernetics Magazine}}} \textbf{\bibinfo{volume}{8}}, \bibinfo{pages}{28--40} (\bibinfo{year}{2022}).

\bibitem{kompa2021second}
\bibinfo{author}{Kompa, B.}, \bibinfo{author}{Snoek, J.} \& \bibinfo{author}{Beam, A.~L.}
\newblock \bibinfo{journal}{\bibinfo{title}{Second opinion needed: communicating uncertainty in medical machine learning}}.
\newblock {\emph{\JournalTitle{NPJ Digital Medicine}}} \textbf{\bibinfo{volume}{4}}, \bibinfo{pages}{4} (\bibinfo{year}{2021}).

\bibitem{luo2023calibrated}
\bibinfo{author}{Luo, Y.}, \bibinfo{author}{Liu, Y.} \& \bibinfo{author}{Peng, J.}
\newblock \bibinfo{journal}{\bibinfo{title}{Calibrated geometric deep learning improves kinase--drug binding predictions}}.
\newblock {\emph{\JournalTitle{Nature Machine Intelligence}}} \bibinfo{pages}{1--12} (\bibinfo{year}{2023}).

\bibitem{olsson2022estimating}
\bibinfo{author}{Olsson, H.} \emph{et~al.}
\newblock \bibinfo{journal}{\bibinfo{title}{Estimating diagnostic uncertainty in artificial intelligence assisted pathology using conformal prediction}}.
\newblock {\emph{\JournalTitle{Nature Communications}}} \textbf{\bibinfo{volume}{13}}, \bibinfo{pages}{7761} (\bibinfo{year}{2022}).

\bibitem{dolezal2022uncertainty}
\bibinfo{author}{Dolezal, J.~M.} \emph{et~al.}
\newblock \bibinfo{journal}{\bibinfo{title}{Uncertainty-informed deep learning models enable high-confidence predictions for digital histopathology}}.
\newblock {\emph{\JournalTitle{Nature Communications}}} \textbf{\bibinfo{volume}{13}}, \bibinfo{pages}{6572} (\bibinfo{year}{2022}).

\bibitem{sun2024tissue}
\bibinfo{author}{Sun, E.~D.}, \bibinfo{author}{Ma, R.}, \bibinfo{author}{Navarro~Negredo, P.}, \bibinfo{author}{Brunet, A.} \& \bibinfo{author}{Zou, J.}
\newblock \bibinfo{journal}{\bibinfo{title}{{TISSUE}: uncertainty-calibrated prediction of single-cell spatial transcriptomics improves downstream analyses}}.
\newblock {\emph{\JournalTitle{Nature Methods}}} \bibinfo{pages}{1--11} (\bibinfo{year}{2024}).

\bibitem{tran2022plex}
\bibinfo{author}{Tran, D.} \emph{et~al.}
\newblock \bibinfo{title}{Plex: Towards reliability using pretrained large model extensions}.
\newblock In \emph{\bibinfo{booktitle}{First Workshop on Pre-training: Perspectives, Pitfalls, and Paths Forward at ICML 2022}} (\bibinfo{year}{2022}).

\bibitem{niculescu2005predicting}
\bibinfo{author}{Niculescu-Mizil, A.} \& \bibinfo{author}{Caruana, R.}
\newblock \bibinfo{title}{Predicting good probabilities with supervised learning}.
\newblock In \emph{\bibinfo{booktitle}{Proceedings of The 22nd International Conference on Machine Learning}}, \bibinfo{pages}{625--632} (\bibinfo{year}{2005}).

\bibitem{kuleshov2018accurate}
\bibinfo{author}{Kuleshov, V.}, \bibinfo{author}{Fenner, N.} \& \bibinfo{author}{Ermon, S.}
\newblock \bibinfo{title}{Accurate uncertainties for deep learning using calibrated regression}.
\newblock In \emph{\bibinfo{booktitle}{International Conference on Machine Learning}}, \bibinfo{pages}{2796--2804} (\bibinfo{organization}{PMLR}, \bibinfo{year}{2018}).

\bibitem{palmer2022calibration}
\bibinfo{author}{Palmer, G.} \emph{et~al.}
\newblock \bibinfo{journal}{\bibinfo{title}{Calibration after bootstrap for accurate uncertainty quantification in regression models}}.
\newblock {\emph{\JournalTitle{NPJ Computational Materials}}} \textbf{\bibinfo{volume}{8}}, \bibinfo{pages}{115} (\bibinfo{year}{2022}).

\bibitem{guo2017calibration}
\bibinfo{author}{Guo, C.}, \bibinfo{author}{Pleiss, G.}, \bibinfo{author}{Sun, Y.} \& \bibinfo{author}{Weinberger, K.~Q.}
\newblock \bibinfo{title}{On calibration of modern neural networks}.
\newblock In \emph{\bibinfo{booktitle}{International Conference on Machine Learning}}, \bibinfo{pages}{1321--1330} (\bibinfo{organization}{PMLR}, \bibinfo{year}{2017}).

\bibitem{ding2021local}
\bibinfo{author}{Ding, Z.}, \bibinfo{author}{Han, X.}, \bibinfo{author}{Liu, P.} \& \bibinfo{author}{Niethammer, M.}
\newblock \bibinfo{title}{Local temperature scaling for probability calibration}.
\newblock In \emph{\bibinfo{booktitle}{Proceedings of the IEEE/CVF International Conference on Computer Vision}}, \bibinfo{pages}{6889--6899} (\bibinfo{year}{2021}).

\bibitem{wilson2020bayesian}
\bibinfo{author}{Wilson, A.~G.} \& \bibinfo{author}{Izmailov, P.}
\newblock \bibinfo{journal}{\bibinfo{title}{Bayesian deep learning and a probabilistic perspective of generalization}}.
\newblock {\emph{\JournalTitle{Advances in Neural Information Processing Systems}}} \textbf{\bibinfo{volume}{33}}, \bibinfo{pages}{4697--4708} (\bibinfo{year}{2020}).

\bibitem{travis2011international}
\bibinfo{author}{Travis, W.~D.} \emph{et~al.}
\newblock \bibinfo{journal}{\bibinfo{title}{International association for the study of lung cancer/american thoracic society/european respiratory society international multidisciplinary classification of lung adenocarcinoma}}.
\newblock {\emph{\JournalTitle{Journal of Thoracic Oncology}}} \textbf{\bibinfo{volume}{6}}, \bibinfo{pages}{244--285} (\bibinfo{year}{2011}).

\bibitem{szegedy2015going}
\bibinfo{author}{Szegedy, C.} \emph{et~al.}
\newblock \bibinfo{title}{Going deeper with convolutions}.
\newblock In \emph{\bibinfo{booktitle}{Proceedings of the IEEE Conference on Computer Vision and Pattern Recognition}}, \bibinfo{pages}{1--9} (\bibinfo{year}{2015}).

\bibitem{kers2022deep}
\bibinfo{author}{Kers, J.} \emph{et~al.}
\newblock \bibinfo{journal}{\bibinfo{title}{Deep learning-based classification of kidney transplant pathology: a retrospective, multicentre, proof-of-concept study}}.
\newblock {\emph{\JournalTitle{The Lancet Digital Health}}} \textbf{\bibinfo{volume}{4}}, \bibinfo{pages}{e18--e26} (\bibinfo{year}{2022}).

\bibitem{chen2024towards}
\bibinfo{author}{Chen, R.~J.} \emph{et~al.}
\newblock \bibinfo{journal}{\bibinfo{title}{Towards a general-purpose foundation model for computational pathology}}.
\newblock {\emph{\JournalTitle{Nature Medicine}}} \bibinfo{pages}{1--13} (\bibinfo{year}{2024}).

\bibitem{lu2024visual}
\bibinfo{author}{Lu, M.~Y.} \emph{et~al.}
\newblock \bibinfo{journal}{\bibinfo{title}{A visual-language foundation model for computational pathology}}.
\newblock {\emph{\JournalTitle{Nature Medicine}}} \textbf{\bibinfo{volume}{30}}, \bibinfo{pages}{863--874} (\bibinfo{year}{2024}).

\bibitem{xu2024whole}
\bibinfo{author}{Xu, H.} \emph{et~al.}
\newblock \bibinfo{journal}{\bibinfo{title}{A whole-slide foundation model for digital pathology from real-world data}}.
\newblock {\emph{\JournalTitle{Nature}}} \bibinfo{pages}{1--8} (\bibinfo{year}{2024}).

\bibitem{ding2024multimodal}
\bibinfo{author}{Ding, T.} \emph{et~al.}
\newblock \bibinfo{journal}{\bibinfo{title}{Multimodal whole slide foundation model for pathology}}.
\newblock {\emph{\JournalTitle{arXiv preprint arXiv:2411.19666}}}  (\bibinfo{year}{2024}).

\bibitem{shafer2008tutorial}
\bibinfo{author}{Shafer, G.} \& \bibinfo{author}{Vovk, V.}
\newblock \bibinfo{journal}{\bibinfo{title}{A tutorial on conformal prediction.}}
\newblock {\emph{\JournalTitle{Journal of Machine Learning Research}}} \textbf{\bibinfo{volume}{9}} (\bibinfo{year}{2008}).

\bibitem{balasubramanian2014conformal}
\bibinfo{author}{Balasubramanian, V.}, \bibinfo{author}{Ho, S.-S.} \& \bibinfo{author}{Vovk, V.}
\newblock \emph{\bibinfo{title}{Conformal prediction for reliable machine learning: theory, adaptations and applications}} (\bibinfo{publisher}{Newnes}, \bibinfo{year}{2014}).

\bibitem{romano2019conformalized}
\bibinfo{author}{Romano, Y.}, \bibinfo{author}{Patterson, E.} \& \bibinfo{author}{Candes, E.}
\newblock \bibinfo{journal}{\bibinfo{title}{Conformalized quantile regression}}.
\newblock {\emph{\JournalTitle{Advances in Neural Information Processing Systems}}} \textbf{\bibinfo{volume}{32}} (\bibinfo{year}{2019}).

\bibitem{coudray2018classification}
\bibinfo{author}{Coudray, N.} \emph{et~al.}
\newblock \bibinfo{journal}{\bibinfo{title}{Classification and mutation prediction from non--small cell lung cancer histopathology images using deep learning}}.
\newblock {\emph{\JournalTitle{Nature Medicine}}} \textbf{\bibinfo{volume}{24}}, \bibinfo{pages}{1559--1567} (\bibinfo{year}{2018}).

\bibitem{campanella2019clinical}
\bibinfo{author}{Campanella, G.} \emph{et~al.}
\newblock \bibinfo{journal}{\bibinfo{title}{Clinical-grade computational pathology using weakly supervised deep learning on whole slide images}}.
\newblock {\emph{\JournalTitle{Nature Medicine}}} \textbf{\bibinfo{volume}{25}}, \bibinfo{pages}{1301--1309} (\bibinfo{year}{2019}).

\bibitem{lu2021data}
\bibinfo{author}{Lu, M.~Y.} \emph{et~al.}
\newblock \bibinfo{journal}{\bibinfo{title}{Data-efficient and weakly supervised computational pathology on whole-slide images}}.
\newblock {\emph{\JournalTitle{Nature Biomedical Engineering}}} \textbf{\bibinfo{volume}{5}}, \bibinfo{pages}{555--570} (\bibinfo{year}{2021}).

\bibitem{chen2021annotation}
\bibinfo{author}{Chen, C.-L.} \emph{et~al.}
\newblock \bibinfo{journal}{\bibinfo{title}{An annotation-free whole-slide training approach to pathological classification of lung cancer types using deep learning}}.
\newblock {\emph{\JournalTitle{Nature Communications}}} \textbf{\bibinfo{volume}{12}}, \bibinfo{pages}{1193} (\bibinfo{year}{2021}).

\bibitem{claudio2024mapping}
\bibinfo{author}{Claudio~Quiros, A.} \emph{et~al.}
\newblock \bibinfo{journal}{\bibinfo{title}{Mapping the landscape of histomorphological cancer phenotypes using self-supervised learning on unannotated pathology slides}}.
\newblock {\emph{\JournalTitle{Nature Communications}}} \textbf{\bibinfo{volume}{15}}, \bibinfo{pages}{4596} (\bibinfo{year}{2024}).

\bibitem{jiang2024transformer}
\bibinfo{author}{Jiang, R.} \emph{et~al.}
\newblock \bibinfo{journal}{\bibinfo{title}{A transformer-based weakly supervised computational pathology method for clinical-grade diagnosis and molecular marker discovery of gliomas}}.
\newblock {\emph{\JournalTitle{Nature Machine Intelligence}}} \textbf{\bibinfo{volume}{6}}, \bibinfo{pages}{876--891} (\bibinfo{year}{2024}).

\bibitem{shahapure2020cluster}
\bibinfo{author}{Shahapure, K.~R.} \& \bibinfo{author}{Nicholas, C.}
\newblock \bibinfo{title}{Cluster quality analysis using silhouette score}.
\newblock In \emph{\bibinfo{booktitle}{2020 IEEE 7th International Conference on Data Science and Advanced Analytics (DSAA)}}, \bibinfo{pages}{747--748} (\bibinfo{organization}{IEEE}, \bibinfo{year}{2020}).

\bibitem{ilse2018attention}
\bibinfo{author}{Ilse, M.}, \bibinfo{author}{Tomczak, J.} \& \bibinfo{author}{Welling, M.}
\newblock \bibinfo{title}{Attention-based deep multiple instance learning}.
\newblock In \emph{\bibinfo{booktitle}{International Conference on Machine Learning}}, \bibinfo{pages}{2127--2136} (\bibinfo{organization}{PMLR}, \bibinfo{year}{2018}).

\bibitem{erickson2020autogluon}
\bibinfo{author}{Erickson, N.} \emph{et~al.}
\newblock \bibinfo{journal}{\bibinfo{title}{Autogluon-tabular: Robust and accurate automl for structured data}}.
\newblock {\emph{\JournalTitle{arXiv preprint arXiv:2003.06505}}}  (\bibinfo{year}{2020}).

\bibitem{cao2023efficient}
\bibinfo{author}{Cao, F.}, \bibinfo{author}{Chen, Q.}, \bibinfo{author}{Xing, Y.} \& \bibinfo{author}{Liang, J.}
\newblock \bibinfo{journal}{\bibinfo{title}{Efficient classification by removing {B}ayesian confusing samples}}.
\newblock {\emph{\JournalTitle{IEEE Transactions on Knowledge and Data Engineering}}} \textbf{\bibinfo{volume}{36}}, \bibinfo{pages}{1084--1098} (\bibinfo{year}{2023}).

\bibitem{liang2022advances}
\bibinfo{author}{Liang, W.} \emph{et~al.}
\newblock \bibinfo{journal}{\bibinfo{title}{Advances, challenges and opportunities in creating data for trustworthy {AI}}}.
\newblock {\emph{\JournalTitle{Nature Machine Intelligence}}} \textbf{\bibinfo{volume}{4}}, \bibinfo{pages}{669--677} (\bibinfo{year}{2022}).

\bibitem{bernhardt2022active}
\bibinfo{author}{Bernhardt, M.} \emph{et~al.}
\newblock \bibinfo{journal}{\bibinfo{title}{Active label cleaning for improved dataset quality under resource constraints}}.
\newblock {\emph{\JournalTitle{Nature Communications}}} \textbf{\bibinfo{volume}{13}}, \bibinfo{pages}{1161} (\bibinfo{year}{2022}).

\bibitem{hager2024evaluation}
\bibinfo{author}{Hager, P.} \emph{et~al.}
\newblock \bibinfo{journal}{\bibinfo{title}{Evaluation and mitigation of the limitations of large language models in clinical decision-making}}.
\newblock {\emph{\JournalTitle{Nature Medicine}}} \bibinfo{pages}{1--10} (\bibinfo{year}{2024}).

\bibitem{pfohl2024toolbox}
\bibinfo{author}{Pfohl, S.~R.} \emph{et~al.}
\newblock \bibinfo{journal}{\bibinfo{title}{A toolbox for surfacing health equity harms and biases in large language models}}.
\newblock {\emph{\JournalTitle{Nature Medicine}}} \bibinfo{pages}{1--11} (\bibinfo{year}{2024}).

\bibitem{lu2023visual}
\bibinfo{author}{Lu, M.~Y.} \emph{et~al.}
\newblock \bibinfo{title}{Visual language pretrained multiple instance zero-shot transfer for histopathology images}.
\newblock In \emph{\bibinfo{booktitle}{Proceedings of the IEEE/CVF Conference on Computer Vision and Pattern Recognition}}, \bibinfo{pages}{19764--19775} (\bibinfo{year}{2023}).

\bibitem{chen2023algorithmic}
\bibinfo{author}{Chen, R.~J.} \emph{et~al.}
\newblock \bibinfo{journal}{\bibinfo{title}{Algorithmic fairness in artificial intelligence for medicine and healthcare}}.
\newblock {\emph{\JournalTitle{Nature Biomedical Engineering}}} \textbf{\bibinfo{volume}{7}}, \bibinfo{pages}{719--742} (\bibinfo{year}{2023}).

\bibitem{oquab2024dinov2}
\bibinfo{author}{Oquab, M.} \emph{et~al.}
\newblock \bibinfo{journal}{\bibinfo{title}{{DINOv2}: Learning robust visual features without supervision}}.
\newblock {\emph{\JournalTitle{Transactions on Machine Learning Research Journal}}} \bibinfo{pages}{1--31} (\bibinfo{year}{2024}).

\bibitem{liu2023simple}
\bibinfo{author}{Liu, J.~Z.} \emph{et~al.}
\newblock \bibinfo{journal}{\bibinfo{title}{A simple approach to improve single-model deep uncertainty via distance-awareness.}}
\newblock {\emph{\JournalTitle{J. Mach. Learn. Res.}}} \textbf{\bibinfo{volume}{24}}, \bibinfo{pages}{42--1} (\bibinfo{year}{2023}).

\bibitem{moor2023foundation}
\bibinfo{author}{Moor, M.} \emph{et~al.}
\newblock \bibinfo{journal}{\bibinfo{title}{Foundation models for generalist medical artificial intelligence}}.
\newblock {\emph{\JournalTitle{Nature}}} \textbf{\bibinfo{volume}{616}}, \bibinfo{pages}{259--265} (\bibinfo{year}{2023}).

\bibitem{huang2023visual}
\bibinfo{author}{Huang, Z.}, \bibinfo{author}{Bianchi, F.}, \bibinfo{author}{Yuksekgonul, M.}, \bibinfo{author}{Montine, T.~J.} \& \bibinfo{author}{Zou, J.}
\newblock \bibinfo{journal}{\bibinfo{title}{A visual--language foundation model for pathology image analysis using medical twitter}}.
\newblock {\emph{\JournalTitle{Nature Medicine}}} \textbf{\bibinfo{volume}{29}}, \bibinfo{pages}{2307--2316} (\bibinfo{year}{2023}).

\bibitem{sun2024pathasst}
\bibinfo{author}{Sun, Y.} \emph{et~al.}
\newblock \bibinfo{title}{Pathasst: A generative foundation ai assistant towards artificial general intelligence of pathology}.
\newblock In \emph{\bibinfo{booktitle}{Proceedings of the AAAI Conference on Artificial Intelligence}}, vol.~\bibinfo{volume}{38}, \bibinfo{pages}{5034--5042} (\bibinfo{year}{2024}).

\bibitem{dolezal2024slideflow}
\bibinfo{author}{Dolezal, J.~M.} \emph{et~al.}
\newblock \bibinfo{journal}{\bibinfo{title}{Slideflow: deep learning for digital histopathology with real-time whole-slide visualization}}.
\newblock {\emph{\JournalTitle{BMC bioinformatics}}} \textbf{\bibinfo{volume}{25}}, \bibinfo{pages}{134} (\bibinfo{year}{2024}).

\bibitem{otsu1975threshold}
\bibinfo{author}{Otsu, N.} \emph{et~al.}
\newblock \bibinfo{journal}{\bibinfo{title}{A threshold selection method from gray-level histograms}}.
\newblock {\emph{\JournalTitle{Automatica}}} \textbf{\bibinfo{volume}{11}}, \bibinfo{pages}{23--27} (\bibinfo{year}{1975}).

\bibitem{liu2020simple}
\bibinfo{author}{Liu, J.} \emph{et~al.}
\newblock \bibinfo{journal}{\bibinfo{title}{Simple and principled uncertainty estimation with deterministic deep learning via distance awareness}}.
\newblock {\emph{\JournalTitle{Advances in Neural Information Processing Systems}}} \textbf{\bibinfo{volume}{33}}, \bibinfo{pages}{7498--7512} (\bibinfo{year}{2020}).

\bibitem{van2020uncertainty}
\bibinfo{author}{Van~Amersfoort, J.}, \bibinfo{author}{Smith, L.}, \bibinfo{author}{Teh, Y.~W.} \& \bibinfo{author}{Gal, Y.}
\newblock \bibinfo{title}{Uncertainty estimation using a single deep deterministic neural network}.
\newblock In \emph{\bibinfo{booktitle}{International Conference on Machine Learning}}, \bibinfo{pages}{9690--9700} (\bibinfo{organization}{PMLR}, \bibinfo{year}{2020}).

\bibitem{o2006metric}
\bibinfo{author}{O'Searcoid, M.}
\newblock \emph{\bibinfo{title}{Metric spaces}} (\bibinfo{publisher}{Springer Science \& Business Media}, \bibinfo{year}{2006}).

\bibitem{behrmann2019invertible}
\bibinfo{author}{Behrmann, J.}, \bibinfo{author}{Grathwohl, W.}, \bibinfo{author}{Chen, R.~T.}, \bibinfo{author}{Duvenaud, D.} \& \bibinfo{author}{Jacobsen, J.-H.}
\newblock \bibinfo{title}{Invertible residual networks}.
\newblock In \emph{\bibinfo{booktitle}{International Conference on Machine Learning}}, \bibinfo{pages}{573--582} (\bibinfo{organization}{PMLR}, \bibinfo{year}{2019}).

\bibitem{gouk2021regularisation}
\bibinfo{author}{Gouk, H.}, \bibinfo{author}{Frank, E.}, \bibinfo{author}{Pfahringer, B.} \& \bibinfo{author}{Cree, M.~J.}
\newblock \bibinfo{journal}{\bibinfo{title}{Regularisation of neural networks by enforcing {L}ipschitz continuity}}.
\newblock {\emph{\JournalTitle{Machine Learning}}} \textbf{\bibinfo{volume}{110}}, \bibinfo{pages}{393--416} (\bibinfo{year}{2021}).

\bibitem{williams2006gaussian}
\bibinfo{author}{Williams, C.~K.} \& \bibinfo{author}{Rasmussen, C.~E.}
\newblock \emph{\bibinfo{title}{Gaussian processes for machine learning}}, vol.~\bibinfo{volume}{2} (\bibinfo{publisher}{MIT press Cambridge, MA}, \bibinfo{year}{2006}).

\bibitem{rahimi2007random}
\bibinfo{author}{Rahimi, A.} \& \bibinfo{author}{Recht, B.}
\newblock \bibinfo{journal}{\bibinfo{title}{Random features for large-scale kernel machines}}.
\newblock {\emph{\JournalTitle{Advances in Neural Information Processing Systems}}} \textbf{\bibinfo{volume}{20}} (\bibinfo{year}{2007}).

\bibitem{angelopoulos2023prediction}
\bibinfo{author}{Angelopoulos, A.~N.}, \bibinfo{author}{Bates, S.}, \bibinfo{author}{Fannjiang, C.}, \bibinfo{author}{Jordan, M.~I.} \& \bibinfo{author}{Zrnic, T.}
\newblock \bibinfo{journal}{\bibinfo{title}{Prediction-powered inference}}.
\newblock {\emph{\JournalTitle{Science}}} \textbf{\bibinfo{volume}{382}}, \bibinfo{pages}{669--674} (\bibinfo{year}{2023}).

\bibitem{lei2018distribution}
\bibinfo{author}{Lei, J.}, \bibinfo{author}{G’Sell, M.}, \bibinfo{author}{Rinaldo, A.}, \bibinfo{author}{Tibshirani, R.~J.} \& \bibinfo{author}{Wasserman, L.}
\newblock \bibinfo{journal}{\bibinfo{title}{Distribution-free predictive inference for regression}}.
\newblock {\emph{\JournalTitle{Journal of the American Statistical Association}}} \textbf{\bibinfo{volume}{113}}, \bibinfo{pages}{1094--1111} (\bibinfo{year}{2018}).

\bibitem{angelopoulos2021gentle}
\bibinfo{author}{Angelopoulos, A.~N.} \& \bibinfo{author}{Bates, S.}
\newblock \bibinfo{journal}{\bibinfo{title}{A gentle introduction to conformal prediction and distribution-free uncertainty quantification}}.
\newblock {\emph{\JournalTitle{arXiv preprint arXiv:2107.07511}}}  (\bibinfo{year}{2021}).

\bibitem{loshchilovdecoupled}
\bibinfo{author}{Loshchilov, I.} \& \bibinfo{author}{Hutter, F.}
\newblock \bibinfo{title}{Decoupled weight decay regularization}.
\newblock In \emph{\bibinfo{booktitle}{International Conference on Learning Representations}} (\bibinfo{year}{2019}).

\bibitem{angelopoulos2022conformal}
\bibinfo{author}{Angelopoulos, A.~N.}, \bibinfo{author}{Bates, S.}, \bibinfo{author}{Fisch, A.}, \bibinfo{author}{Lei, L.} \& \bibinfo{author}{Schuster, T.}
\newblock \bibinfo{journal}{\bibinfo{title}{Conformal risk control}}.
\newblock {\emph{\JournalTitle{arXiv preprint arXiv:2208.02814}}}  (\bibinfo{year}{2022}).

\bibitem{ktena2024generative}
\bibinfo{author}{Ktena, I.} \emph{et~al.}
\newblock \bibinfo{journal}{\bibinfo{title}{Generative models improve fairness of medical classifiers under distribution shifts}}.
\newblock {\emph{\JournalTitle{Nature Medicine}}} \bibinfo{pages}{1--8} (\bibinfo{year}{2024}).

\end{thebibliography}

\clearpage

\captionsetup[figure]{labelformat=empty}
\setcounter{figure}{0}

\captionsetup[table]{labelformat=empty}

\begin{figure}[!ht]
    \centering
    \includegraphics[scale=0.210]{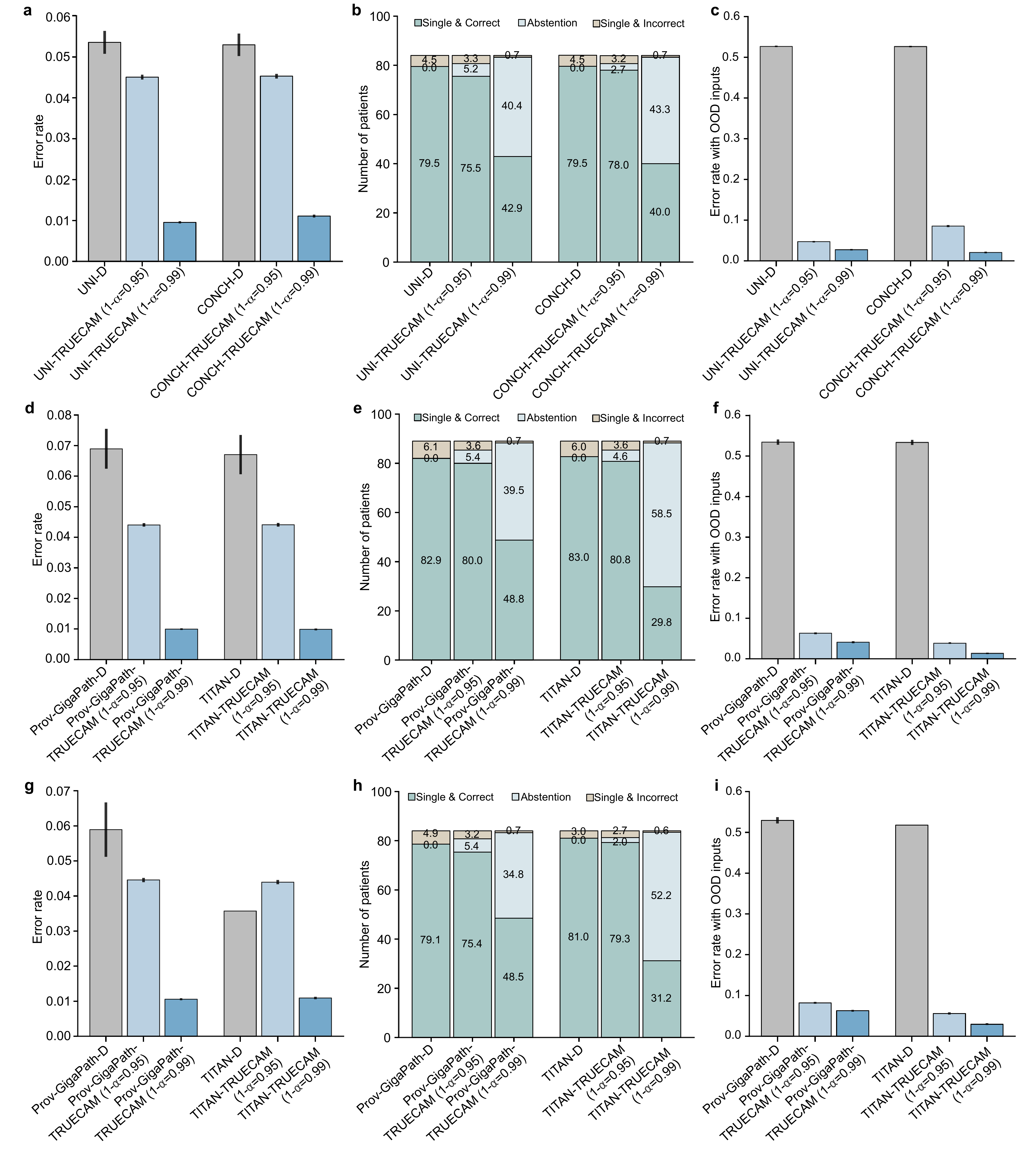}
    \caption{\textbf{Extended Data Figure 1}: Evaluation of TRUECAM's performance across four foundation models: UNI, CONCH, Prov-GigaPath, and TITAN. 
    \textbf{a, d, g}, TRUECAM significantly reduced NSCLC subtyping error rates across model types (denoted using suffix ``-TRUECAM'') compared to their original deterministic versions (denoted using suffix ``-D''), adhering to the pre-specified true label coverage  1-$\alpha$. The only exception is TITAN-TRUECAM with $1-\alpha=0.95$, where TITAN-D achieved an error rate below 0.05. As a result, CP produced a zero prediction set size (no prediction) to maintain the desired coverage of 0.95. \textbf{b, e, h}, Patient-level classification breakdown for models with and without TRUECAM. \textbf{c, f, i}, TRUECAM achieved significantly lower error rates in real-world NSCLC subtyping scenarios involving a 1:1 mix of in-domain and out-of-domain inputs. Evaluations in \textbf{a}-\textbf{c} and \textbf{g}-\textbf{i} are based on the CPTAC testing data, whereas those in \textbf{d-f} are based on the TCGA testing data. Results shown in \textbf{a}-\textbf{c} are based on 20 independently trained models, each with 500 conformal prediction evaluations, and those in \textbf{d}-\textbf{i} are based on 20 independently trained models. OOD, out-of-domain; D, Deterministic.}  
    \label{fig:performance-summary-cptac}
\end{figure}

\begin{figure}[!ht]
    \centering
    \includegraphics[scale=0.215]{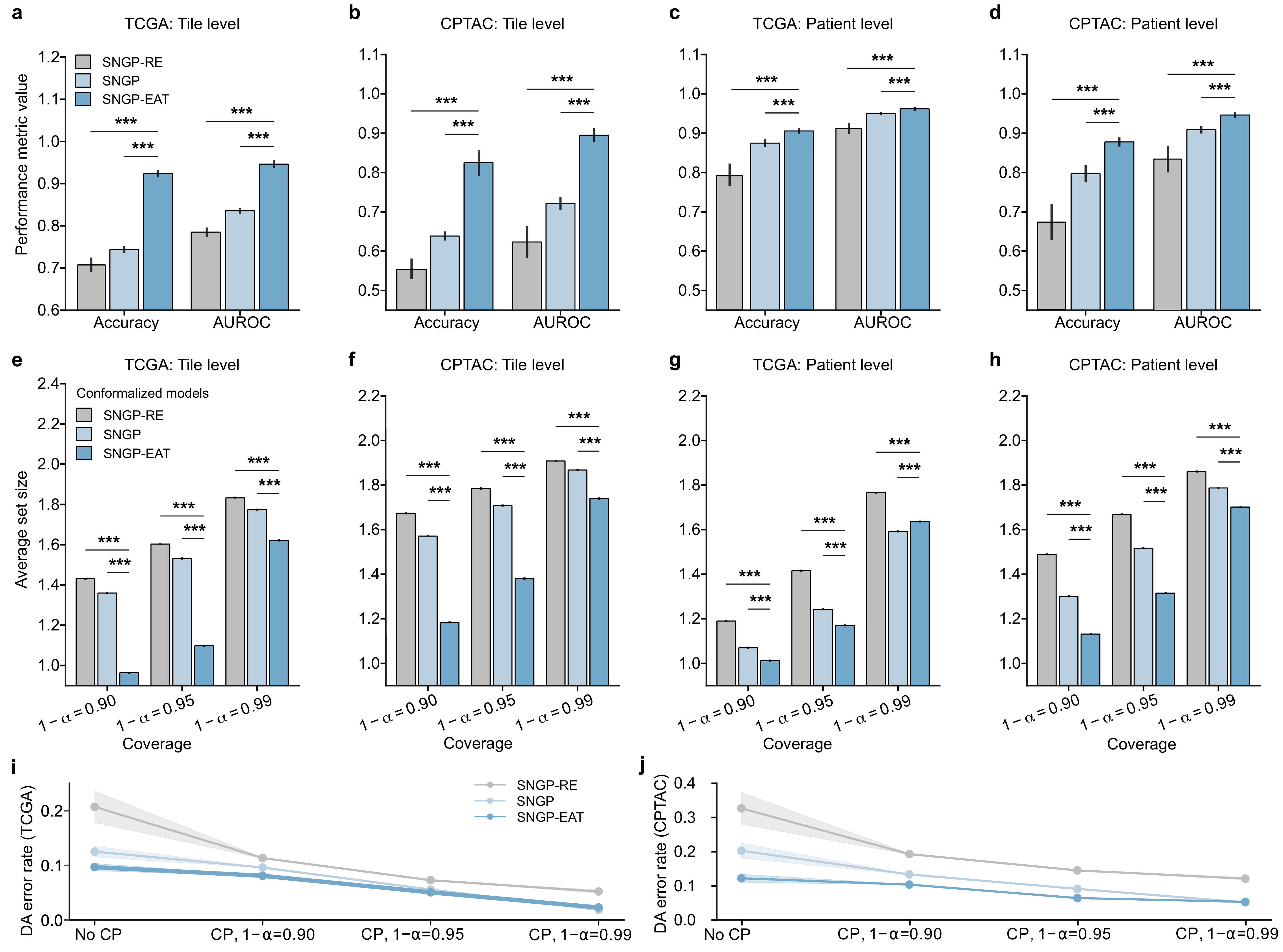}
    \caption{\textbf{Extended Data Figure 2}: Impact assessment of EAT on classification and CP performance. \textbf{a}, Tile-level classification performance on TCGA. \textbf{b}, Tile-level classification performance on CPTAC. \textbf{c}, Patient-level classification performance on TCGA. \textbf{d}, Patient-level classification performance on CPTAC. \textbf{e}, Tile-level prediction set size on TCGA for three distinct values of significance level $\alpha$. \textbf{f}, Tile-level prediction set size on CPTAC. \textbf{g}, Patient-level prediction set size on TCGA. \textbf{h}, Patient-level prediction set size on CPTAC. \textbf{i}, Patient-level DA error rate on TCGA before and after activating CP. \textbf{j}, Patient-level DA error rate on CPTAC before and after activating CP. Mean values and the corresponding 95\% confidence intervals in \textbf{a}-\textbf{j} are based on 20 independently trained models, each with 500 CP evaluations (See details in the Methods). One-sided Wilcoxon tests are utilized to calculate \textit{p} values. *** $p<0.001$. SNGP, spectral-normalized neural Gaussian process; RE, random elimination; EAT, elimination of ambiguous tiles; DA, definitive-answer.}
    \label{fig:extended_data_figure1}
\end{figure}

\begin{figure}
    \centering\includegraphics[scale=0.21]{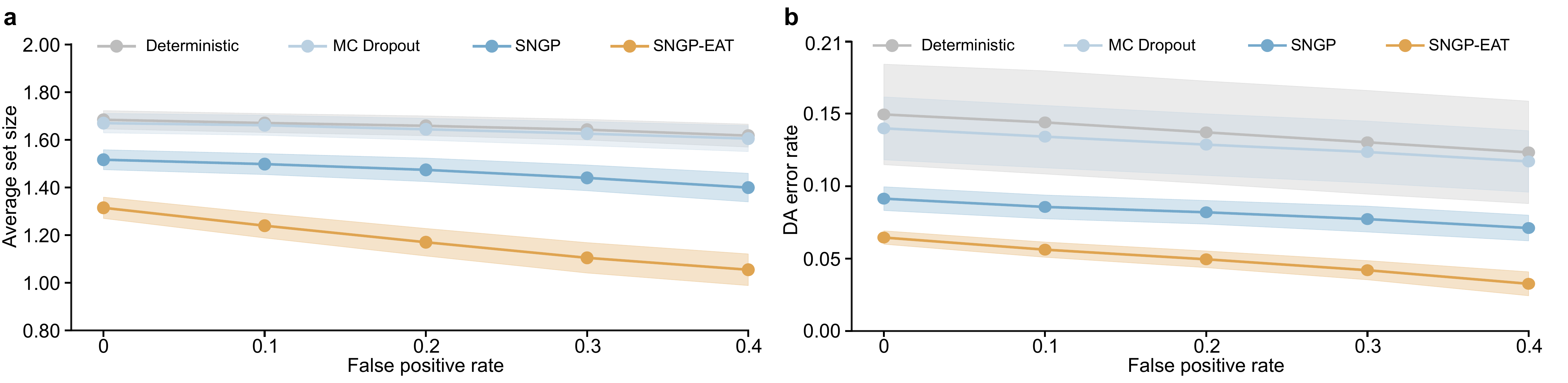}
    \caption{\textbf{Extended Data Figure 3}. Impact of OOD detection-based distribution shift control on the average set size of CP and the definitive-answer error rate in the CPTAC testing dataset. \textbf{a}, Average set size as a function of false positive rate for OOD detection after applying distribution shift control. \textbf{b}, Definitive-answer error rate as a function of false positive rate for OOD detection after applying distribution shift control. Mean values and the corresponding 95\% confidence intervals are based on 20 independently trained models, each with 500 CP evaluations (See details in the Methods). SNGP, spectral-normalized neural Gaussian process; EAT, elimination of ambiguous tiles; DA, definitive-answer.}
    \label{fig:SNGP_CP_complementary_summary_extended}
\end{figure}

\begin{figure}
    \centering
    \includegraphics[scale=0.24]{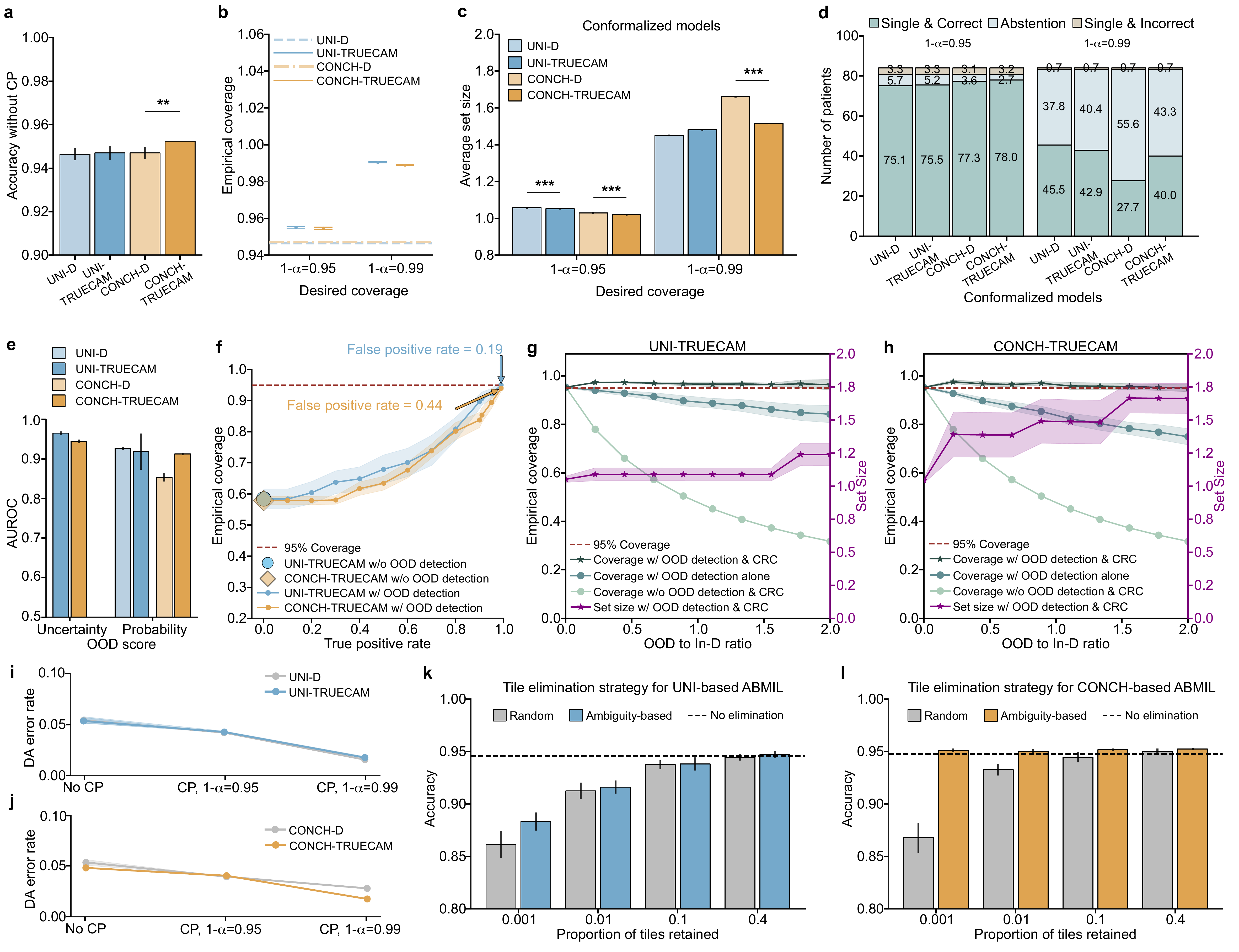}
    \caption{\textbf{Extended Data Figure 4}. Assessment of TRUECAM using the foundation models UNI and CONCH. Results on CPTAC are reported. \textbf{a}, Classification accuracy of foundation model-based ABMIL with (denoted using suffix ``-TRUECAM'') and without TRUECAM (i.e., the original version of foundation model-based ABMIL, denoted using suffix ``-D''), evaluated without CP. \textbf{b}, Comparison of empirical coverage with and without TRUECAM. \textbf{c}, Comparison of average set size of CP with and without TRUECAM. \textbf{d}, Comparison of patient classification breakdown with and without TRUECAM. \textbf{e}, Performance of two distinct OOD scores in identifying slides outside a model’s scope, which was established using the CPTAC training data. The uncertainty-based score is not applicable to the original versions of UNI and CONCH. \textbf{f}, Empirical coverage with $\alpha$ = 0.05 as a function of the true positive rate for OOD detection. \textbf{g}, \textbf{h}, Empirical coverage and set size of CP for UNI-TRUECAM and CONCH-TRUECAM under conformal risk control with a false positive rate of 0.19 for UNI and 0.44 for CONCH, shown as a function of OOD-to-In-D ratio. \textbf{i}, \textbf{j}, Comparison of patient-level DA error rate with and without TRUECAM in varying contexts of CP. \textbf{k}, \textbf{l}, Comparison of classification accuracy for UNI-TRUECAM and CONCH-TRUECAM using TRUECAM's ambiguity-based tile elimination strategy versus random tile elimination, shown as a function of the proportion of tiles retained per slide. One-sided Wilcoxon signed-rank test is utilized to calculate \textit{p} values. ** $p<0.01$, *** $p<0.001$. ABMIL, attention-based multiple instance learning; CP, conformal prediction; SNGP, spectral-normalized neural Gaussian process; EAT, elimination of ambiguous tiles; DA, definitive-answer.}
    \label{fig:foundation_model_CPTAC_summary}
\end{figure}

\begin{figure}[!ht]
    \centering
    \includegraphics[scale=0.21]{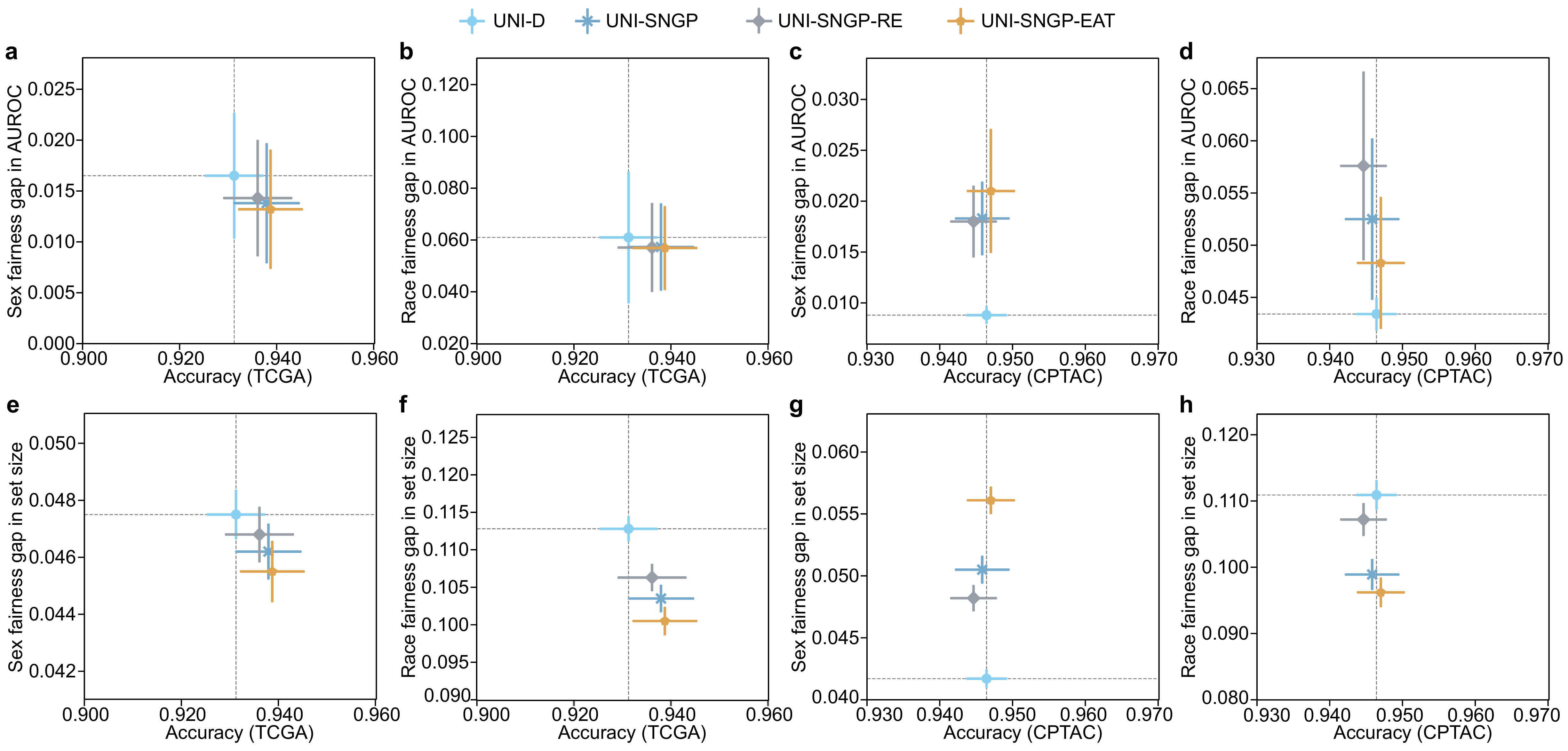}
    \caption{\textbf{Extended Data Figure 5}: Fairness evaluation of UNI-based ABMIL, measured by the metric value differences between the best- and worst-performing demographic subgroups against the overall accuracy of NSCLC subtyping. \textbf{a-d}, Sex- and race-wise AUROC gap versus overall accuracy for discriminating between LUAD and LUSC on TCGA (n=89) and CPTAC (n=84) with CP deactivated. \textbf{e-h}, At $\alpha = 0.05$, sex- and race-wise set size gap versus overall accuracy on TCGA and CPTAC with CP activated. Evaluations on UNI-D and its conformalized version are marked as an overall baseline.}
    \label{fig:extended_figure_UNI_fairness}
\end{figure}


\begin{figure}[!ht]
    \centering
    \includegraphics[scale=0.21]{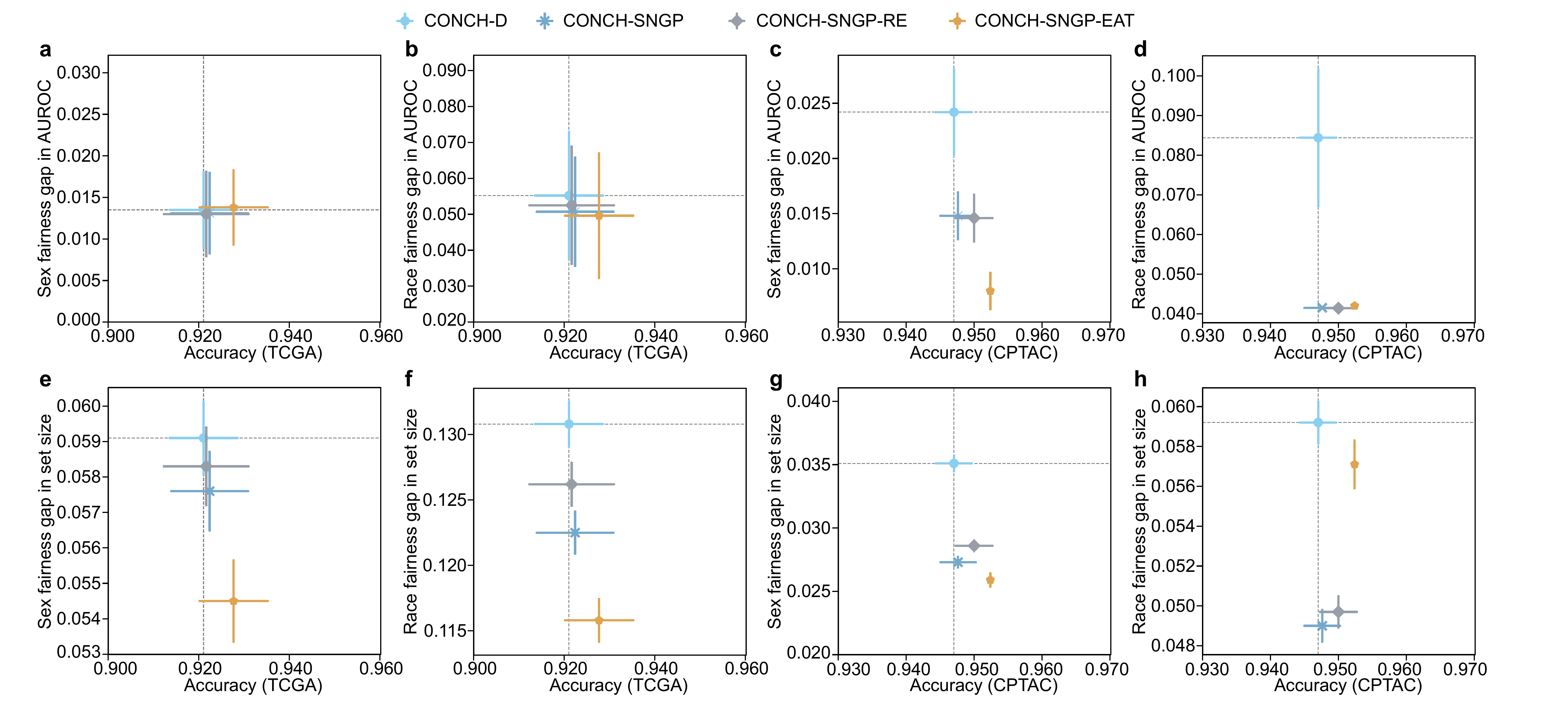}
    \caption{\textbf{Extended Data Figure 6}: Fairness evaluation of CONCH-based ABMIL, measured by the metric value differences between the best- and worst-performing demographic subgroups against the overall accuracy of NSCLC subtyping. \textbf{a-d}, Sex- and race-wise AUROC gap versus overall accuracy for discriminating between LUAD and LUSC on TCGA (n=89) and CPTAC (n=84) with CP deactivated. \textbf{e-h}, At $\alpha = 0.05$, sex- and race-wise set size gap versus overall accuracy on TCGA and CPTAC with CP activated. Evaluations on CONCH-D and its conformalized version are marked as an overall baseline.}
    \label{fig:extended_figure_CONCH_fairness}
\end{figure}

\begin{figure}[!ht]
    \centering
    \includegraphics[scale=0.29]{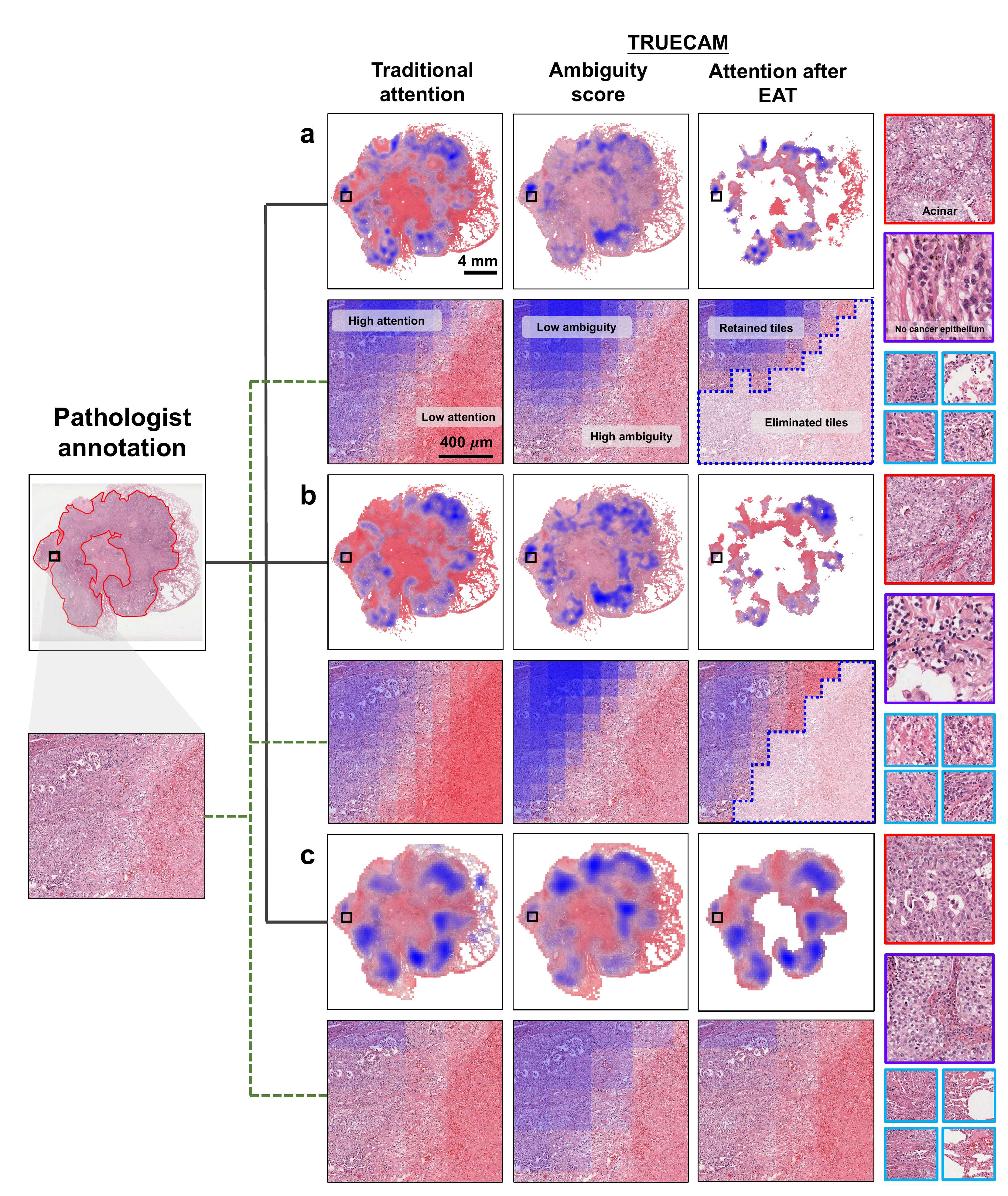}
    \caption{\textbf{Extended Data Figure 7}: Visualization of a correctly classified LUAD slide randomly selected from the TCGA testing dataset, accompanied by a pathologist's annotations and model-derived information supporting the inference. The whole-slide image (WSI) and a selected patch (with a zoomed-in view) are visualized in the context of \textbf{a}, UNI, \textbf{b}, CONCH, and \textbf{c}, Inception-v3. The red-highlighted region in the original WSI (first column) indicates the pathologist’s annotation used to guide NSCLC subtyping. Traditional interpretability (second column) was generated from the attention weights of the ABMIL module in foundation models or from the predicted tile-level subtype probabilities in Inception-v3. Tile-level ambiguity scores (third column) were derived from the UQ module or UQ-informed subtype probabilities in TRUECAM. After applying ambiguity-guided EAT, the attention map (fourth column) was calculated for inference using the SNGP-EAT version of each model with white regions indicating the eliminated tiles. Example tiles (all from the selected patch) visualized in the fifth column include: a randomly selected tile with low ambiguity and a high-attention weight (red border), a tile with low ambiguity and a low-attention weight (purple border), and four randomly selected tiles with high ambiguity (removed by EAT, blue border). The pathologist annotated these tiles as ``good indication'', ``cannot determine'', and ``not relevant'', respectively, demonstrating strong alignment with the TRUECAM-derived interpretability.}
    \label{fig:first_concatenated}
\end{figure}

\begin{figure}[!ht]
    \centering
    \includegraphics[scale=0.29]{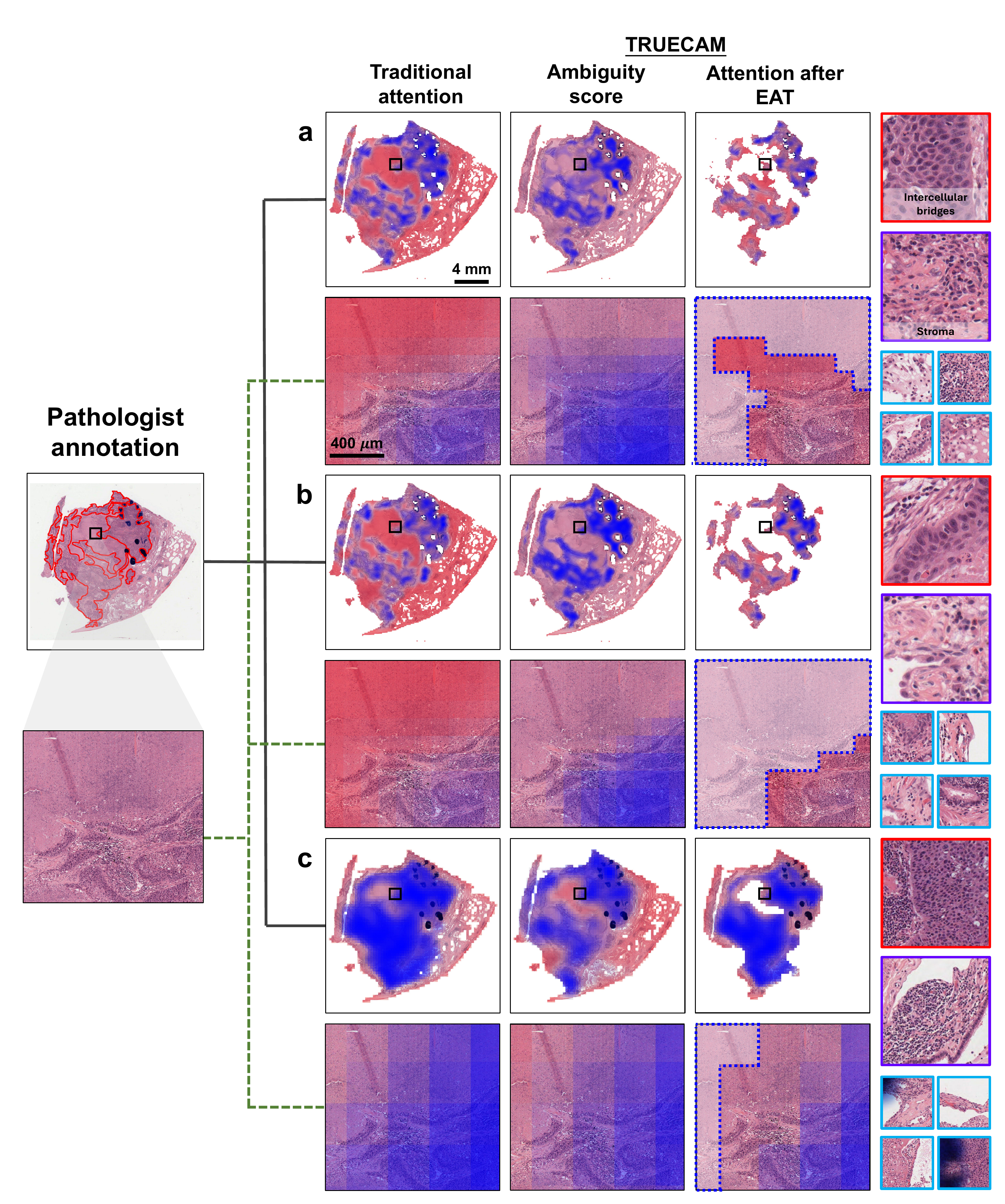}
    \caption{\textbf{Extended Data Figure 8}: Visualization of a correctly classified LUSC slide randomly selected from the TCGA testing dataset, accompanied by a pathologist's annotations and model-derived information supporting the inference. The whole-slide image (WSI) and a selected patch (with a zoomed-in view) are visualized in the context of \textbf{a}, UNI, \textbf{b}, CONCH, and \textbf{c}, Inception-v3. The red-highlighted region in the original WSI (first column) indicates the pathologist’s annotation used to guide NSCLC subtyping. Traditional interpretability (second column) was generated from the attention weights of the ABMIL module in foundation models or from the predicted tile-level subtype probabilities in Inception-v3. Tile-level ambiguity scores (third column) were derived from the UQ module or UQ-informed subtype probabilities in TRUECAM. After applying ambiguity-guided EAT, the attention map (fourth column) was calculated for inference using the SNGP-EAT version of each model with white regions indicating the eliminated tiles. Example tiles (all from the selected patch) visualized in the fifth column include: a randomly selected tile with low ambiguity and a high-attention weight (red border), a tile with low ambiguity and a low-attention weight (purple border), and four randomly selected tiles with high ambiguity (removed by EAT, blue border). The pathologist annotated these tiles as ``good indication'', ``cannot determine'', and ``not relevant'', respectively, demonstrating strong alignment with the TRUECAM-derived interpretability.}
    \label{fig:second_concatenated}
\end{figure}

\captionsetup[figure]{labelformat=empty}
\begin{figure}[!ht]
    \centering
    \includegraphics[scale=0.22]{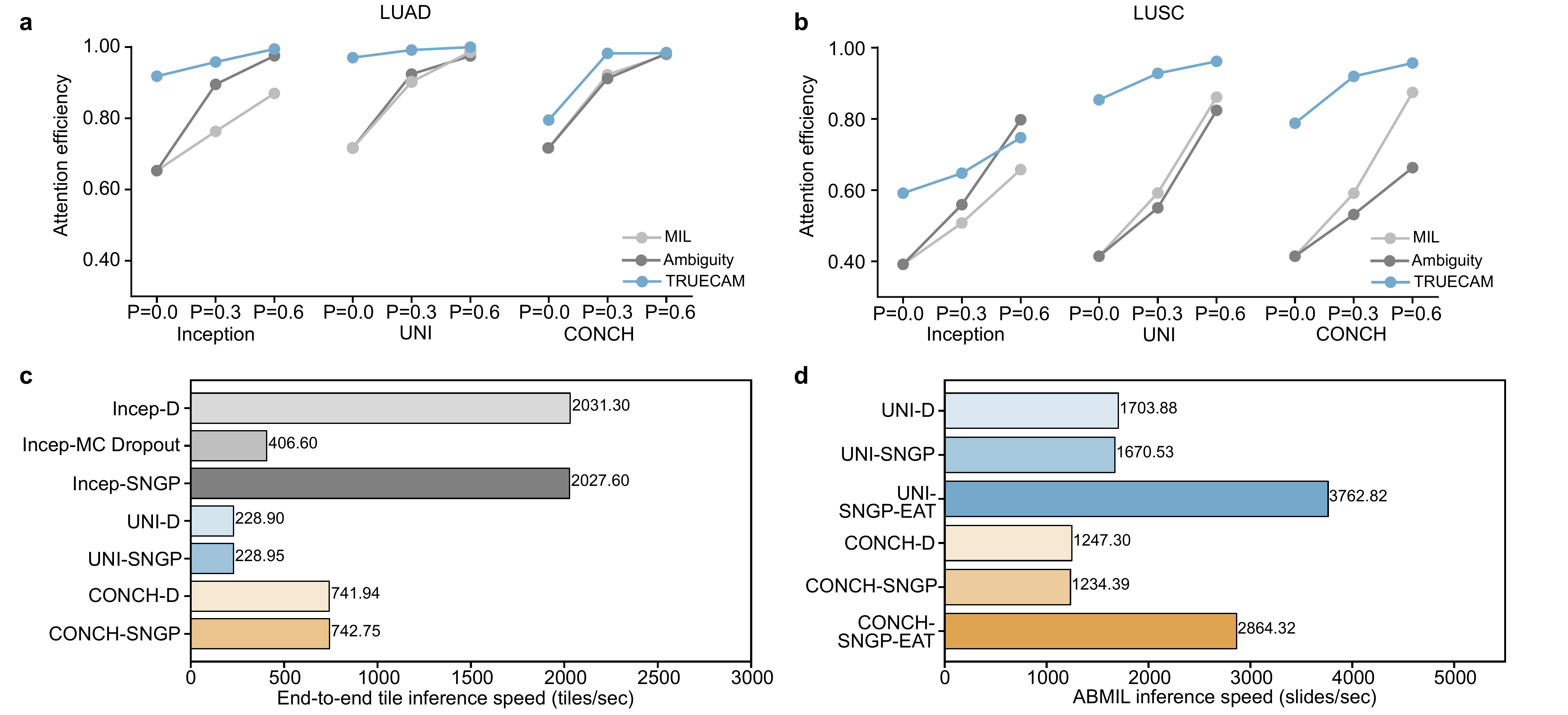}
    \caption{\textbf{Extended Data Figure 9}: Comparison of attention efficiency and  inference speed across different model settings. \textbf{a}, Comparison of attention efficiency for traditional attention (i.e., MIL), ambiguity score, and TRUECAM (i.e., attention after EAT) with respect to the LUAD slide shown in Extended Data Fig.~\ref{fig:first_concatenated}. $P$ represents the proportion of tiles with the lowest attention scores removed. In the ``MIL'' setting, attention scores for Inception-v3 and two foundation models correspond to the tile probability and corresponding ABMIL attention scores, respectively, whereas scores in the setting of ``Ambiguity'' for Inception-v3 and two foundation models correspond to the ambiguity scores from their own models and AutoGluon models, respectively. \textbf{b}, Comparison of attention efficiency for traditional attention (i.e., MIL), ambiguity score, and TRUECAM (i.e., attention after EAT) with respect to the LUAD slide shown in Extended Data Fig.~\ref{fig:second_concatenated}. \textbf{c}, Comparison of end-to-end tile inference speed. \textbf{d}, Comparison of slide-level inference speed for ABMIL-based classification models using pre-extracted tile-level representations from foundation models. 
    Inference speed was measured using an NVIDIA GeForce 4090 (24GB) with a batch size of 512 and 32-bit floating-point precision. Benchmarking excluded the preprocessing and load time with 10 warm-up iterations followed by 30 benchmark iterations. D, Deterministic; MIL, multiple instance learning; ABMIL, attention-based multiple instance learning; SNGP, spectral-normalized neural Gaussian process; EAT, elimination of ambiguous tiles.}
    \label{fig:attention_inference_efficiency}
\end{figure}


\end{document}